\documentclass[review, 12pt, a4paper]{elsarticle}

\usepackage{setspace}
\usepackage{lipsum}
\usepackage{makeidx}
\usepackage{subcaption}
\usepackage{enumerate}
\usepackage{color}

\usepackage{amsmath}
\usepackage{amssymb}
\usepackage{amsthm}
\usepackage{nomencl}
\usepackage{multirow}
\usepackage{graphicx}
\usepackage{anysize}
\usepackage{float}
\usepackage{epstopdf}
\usepackage{threeparttable}
\usepackage{multicol}
\usepackage[table]{xcolor}
\usepackage[font=footnotesize]{caption}

\DeclareGraphicsExtensions{.pdf,.jpeg,.png,.jpg,.emf,.eps}
\usepackage{geometry}
\usepackage{hyperref}

\journal{Nano Communication Networks}

\usepackage{fancyhdr}
\makeatletter
\def\ps@pprintTitle{%
	\let\@oddhead\@empty
	\let\@evenhead\@empty
	\def\@oddfoot{\parbox{\textwidth}{\centering\scriptsize THIS IS AN AUTHOR-CREATED POSTPRINT VERSION Submitted to Nano communication networks\\[1ex]\thepage}}%
	\let\@evenfoot\@oddfoot
}
\makeatother

\begin{document}
	
	\begin{frontmatter} 
		
		\title{Array-Based Molecular Pulse Encoding for Neuro-Spike Communication in Intra-Body Nano-networks\vspace*{-0.3em}}
		
		\author{Keyvan Aghababaiyan}
		\ead{kaghababaiyan@umh.es}
		\affiliation{organization={Universidad Miguel Hernández de Elche},
			addressline={03202 Elche},
			country={Spain}}
		
		\begin{abstract}
			In this paper, we investigate a neuro-spike communication system designed to bridge severed connections between damaged neurons using auxiliary nano-machines. Natural neuro-spike communication typically relies on instantaneous spike rates and temporal intervals to convey information. However, these temporal encoding schemes require exact time synchronization between the transmitter and receiver, a requirement that poses a significant challenge for resource-constrained nano-machines. To address this issue, it is imperative for future intra-body nano-networks to develop communication schemes that operate under reduced-order synchronization (e.g., symbol-synchronized). In this paper, we propose a novel neuro-spike array-based communication scheme where information is encoded through the specific arrangement of distinct molecular pulses emitted by nano-machines. By distinguishing symbols based on the sequence of these emissions rather than their exact timing, the need for stringent time synchronization is eliminated. We theoretically analyze the performance of the proposed scheme by deriving expressions for the probability of inter-symbol interference (ISI), error probability, and the achievable communication rate. Analytical and numerical results demonstrate that our array-based scheme significantly outperforms previously proposed symbol-synchronized models, providing a $75\% - 150\%$ enhancement in the communication rate across various diffusion coefficients.
		\end{abstract}
		
		
	\end{frontmatter} 
	
	\section{Introduction}
	
	Recent advancements in nanotechnology and communication engineering have catalyzed the development of a new generation of nanoscale devices designed for intra-body integration \cite{jornet2013graphene}. These nano-machines, despite their limited individual resources, can coordinate within a nanonetwork to perform sophisticated medical tasks, ranging from real-time diagnostics to targeted therapeutic interventions. A particularly vital frontier is the deployment of these nanonetworks within neuronal tissue to monitor or stimulate damaged nervous systems \cite{seo2013neural, suzuki2014service}. By establishing a bridge between healthy and impaired neural regions, coordinated nano-machines offer a promising solution to restore severed connections and recover lost physiological functions \cite{mesiti2013nanomachine}.
	
	To achieve such seamless integration, communication techniques must be inspired by the biological principles of the nervous system. Neuro-spike communication represents the primary hybrid mechanism for information transfer in this domain, involving the diffusion of neurotransmitters across the synaptic cleft and the subsequent propagation of action potentials along the axonal pathway. In natural systems, neural coding determines how stimuli are mapped onto these electrical impulses \cite{dayan2001theoretical}. Traditionally, neurons utilize spike-rate and temporal coding, where information is embedded in the frequency of spikes or the precise timing between them. However, a major bottleneck for artificial nano-machines is that these coding schemes necessitate near-perfect time synchronization between the transmitter and the receiver. Without exact knowledge of spike arrival times, the reliability of the decoded information deteriorates significantly.
	
	Over the past decade, extensive research has characterized the neuro-spike channel from various perspectives. Initial efforts focused on stochastic modeling of signal processing—such as adaptive threshold models for spike probability \cite{ref201010, jadid3}—and physical representations of hippocampal communication \cite{ref8, jadid4}. As the field matured, the analysis expanded to include specific synaptic behaviors, such as neurotransmitter binding kinetics \cite{khan2017diffusion, ref9}, upper bounds on synaptic capacity \cite{ref201616}, and realistic pool-based vesicle release mechanisms \cite{ramezani2018information, yu2022modeling}. More recently, time-slotted channel frameworks with relay synaptic neurons have been proposed to mitigate inter-symbol interference (ISI) and improve the maximum achievable rate in synaptic gaps \cite{chen2023bio}. In parallel, the impact of axonal noise and variability on signal transmission was rigorously investigated to understand the limits of axonal pathways \cite{k1}. Furthermore, accurate error probability modeling has been developed \cite{ref4, aghababaiyan2018axonal}. Other related factors include synaptic noise \cite{k2, s13534_2024}.
	
	In recent years, the scope of neuro-spike research has evolved to incorporate more complex biological dynamics and broader network architectures, particularly within the Internet of Nano-Things (IoNT). For instance, recent experimental platforms using biological nerve-muscle channels have validated the feasibility of through-body neural communication \cite{s41467_2024}. To address signal degradation in such environments, advanced models have been proposed to analyze both external and internal interference \cite{s13534_2024}, and novel coding schemes have been developed that leverage the subthreshold oscillatory characteristics of membrane potentials for improved reliability \cite{ioNT_membrane202X}. Furthermore, recent studies have emphasized the role of synaptic plasticity, demonstrating how Spike-Time Dependent Plasticity (STDP) and memory consolidation significantly influence the mutual information and adaptive strength of neural links \cite{stdp_impact202X, stdp_mutual202X}. This line of research highlights the increasingly sophisticated modeling of the neural communication environment.
	
	Despite these advancements, a critical challenge remains: traditional neural encoding paradigms, such as rate coding and temporal coding, heavily depend on synchronized time windows or exact spike instants, demanding precise, high-order synchronization. While certain approaches have sought to relax these strict constraints by utilizing symbol-synchronized models like On-Off Keying (OOK), they still strictly require aligned time slots to function and inherently face timing uncertainties and jitter \cite{aghababaiyan2018asynchronous}.
	
	In this paper, we propose a novel neuro-spike array-based communication scheme designed to circumvent the synchronization barrier. Inspired by the structural arrangement of information in DNA arrays, our scheme encodes information in the sequential arrangement of distinct molecular pulses (e.g., using different neurotransmitter types). The term "array-based" is adopted to emphasize that information is mapped onto a strictly ordered sequence (an array) of structurally distinct molecular elements. Rather than relying on a single type of molecule varying over time, the transmitter constructs a temporal array of different neurotransmitters to convey symbols. By utilizing specific target receptors for signal distinction, the receiving nano-machine (RN) can decode the information without needing to track the exact arrival times of individual pulses. This approach not only simplifies the synchronization requirements but also provides a significantly higher communication rate compared to traditional symbol-synchronized models.
	
	The main difference between our proposed architecture and existing neuro-spike models is the elimination of strict time-synchronization requirements. While conventional methods depend on precise instantaneous timing or exact spike intervals, our scheme decodes information purely based on the sequential chemical arrangement of distinct pulses, thereby offering a highly robust mechanism that significantly outperforms traditional models in terms of channel capacity.
	
	The key contributions of this paper are:
	\begin{itemize}
		\item We propose an array-based communication architecture for neuro-spike links where information is encoded through the arrangement of distinct molecular pulses, modeled under the influence of additive Gamma noise \cite{ref999}.
		\item We derive analytical expressions for the probability of inter-symbol interference (ISI) and the bit error probability, providing a theoretical basis for calculating the achievable communication rate.
		\item Through numerical evaluation, we demonstrate that our scheme yields a $75\% - 150\%$ enhancement in communication rate compared to conventional on/off keying (OOK) in symbol-synchronized channels.
	\end{itemize}
	
	The rest of the paper is organized as follows. Section 2 describes the system model. Section 3 presents the mathematical analysis of the proposed array-based scheme. Numerical results are discussed in Section 4. Section 5 provides a discussion on the practical limitations and biological constraints of the model. Finally, Section 6 concludes the paper.
	
	\section{System Model}
	
	Neurodegenerative diseases are debilitating conditions characterized by the progressive degeneration or death of nerve cells, ultimately leading to a loss of neuronal connectivity. Although the typical soma size of these neurons ranges from 4 to 100 $\mu\text{m}$, from a molecular communication perspective, the synaptic cleft provides sufficient spatial capacity for the deployment of nanoscale devices. Consequently, auxiliary nano-machines can be employed to bypass damaged pathways and seamlessly reconnect severed neural links.
	
	Fig. \ref{picc-1} illustrates a scenario where nano-machines restore communication between impaired neurons. Specifically, in conditions such as Amyotrophic Lateral Sclerosis (ALS) or severe traumatic nerve injuries, the physical axonal pathway may be completely severed, or the presynaptic terminals may necrotize. In such cases, the natural release of neurotransmitters (e.g., Glutamate and GABA) ceases entirely, or the molecules fail to traverse the degraded synaptic cleft to reach the post-synaptic receptors. If the pre-synaptic neuron is damaged, pre-synaptic nano-machines (acting as transmitters) release neurotransmitters into the synaptic cleft upon receiving external stimuli. Conversely, if the post-synaptic neuron is unresponsive, post-synaptic nano-machines (acting as receivers) detect neurotransmitters in the cleft and subsequently induce an artificial stimulus into the post-synaptic neuron.
	
	In this study, we model the communication between neurons located in the motor cortex—a region of the cerebral cortex responsible for the planning, control, and execution of voluntary movements. Furthermore, we assume that information is conveyed using temporal modulation. The fundamental components and assumptions of our system model are detailed as follows:
	
	\begin{figure}[t]
		\centering
		\includegraphics[width=.68\textwidth]{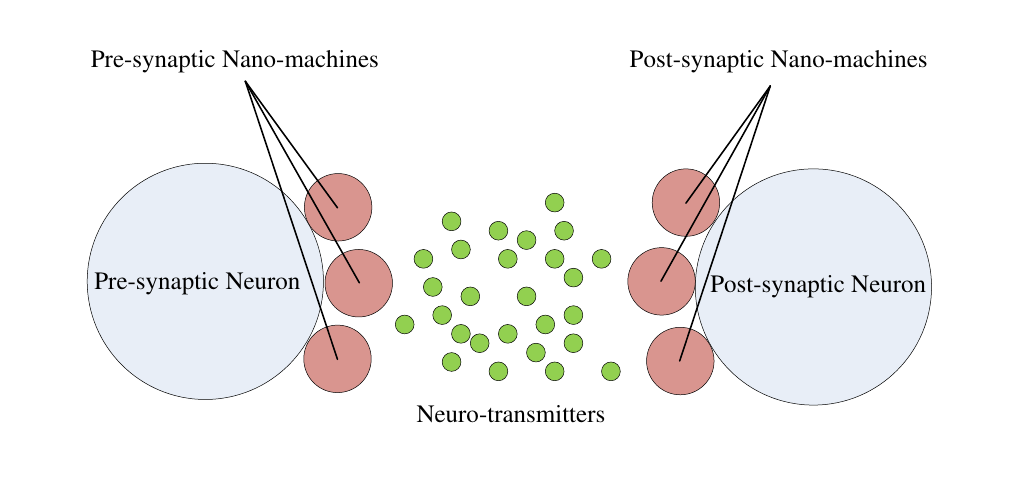}
		\caption{Employing nano-machines for stimulation tasks when neurons have lost their ability to communicate.}
		\label{picc-1}
	\end{figure}
	
	\subsection{Synaptic Transmission}
	In natural chemical synapses, cells communicate by releasing neurotransmitters that traverse the synaptic cleft and bind to specific receptors. This binding alters the ion flow across the post-synaptic membrane, producing either excitatory (increasing the probability of generating an action potential) or inhibitory (decreasing the probability) effects. In our model, we exploit two primary types of neurotransmitter molecules: Glutamate and Gamma-Aminobutyric Acid (GABA). Glutamate serves as the principal fast excitatory neurotransmitter in the brain, while GABA acts as the primary fast inhibitory neurotransmitter \cite{petroff2002book}. By utilizing these two distinct molecular types, the nano-machines can encode and transmit binary information reliably \cite{aghababaiyan2018joint}.
	
	\subsection{Nanomachine Logic and Pulse Generation}
\begin{figure}[t]
		\centering
		\includegraphics[width=.68\textwidth]{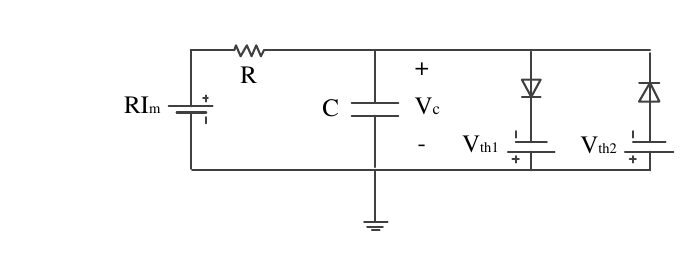}
		\caption{The internal logic circuit of the transmitting nano-machine, based on a modified I\&F model with dual thresholds.}
		\label{picc-2}
	\end{figure}

	To govern the emission of these neurotransmitters, we assume the transmitting nano-machine operates based on a modified, dual-threshold Integrate-and-Fire (I\&F) logic circuit, as depicted in Fig. \ref{picc-2} \cite{fourcaud2002dynamics}. Unlike a natural neuron, which fires a uniform action potential (following the all-or-none principle), this internal logic circuit incorporates two distinct voltage thresholds, $V_{\text{th}1}$ and $V_{\text{th}2}$, designed to trigger the release of different molecular pulses. 
	
	We assume the potential differences between the initial resting voltage of the circuit's capacitance, $V_c(0)$, and the two thresholds are symmetric in magnitude but opposite in polarity; i.e., $\left| {{V_c}(0) - {V_{\text{th}1}}} \right| = \left| {{V_c}(0) - {V_{\text{th}2}}} \right|$. Within this extended I\&F framework, accumulating excitatory stimuli causes depolarization (adding positive charge to the capacitance), whereas inhibitory stimuli induce hyperpolarization (adding negative charge).
	
	Based on this logic, the auxiliary nano-machine can generate two distinct types of molecular emissions, referred to as pulses rather than classical electrical spikes. A pulse of type $A$ (e.g., a burst of Glutamate) is emitted whenever the internal voltage ${V_c}(t)$ exceeds the upper threshold $V_{\text{th}1}$. Conversely, a pulse of type $B$ (e.g., a burst of GABA) is emitted whenever ${V_c}(t)$ drops below the lower threshold $V_{\text{th}2}$. Because the receiving nano-machine is equipped with specialized molecular receptors, it can easily distinguish between pulse $A$ and pulse $B$ chemically, entirely eliminating the need to discriminate signals based on amplitude variations.
	
	\subsection{Modeling the Jitter of Propagation Time as Channel Noise}
	In a neuro-spike communication channel that exploits temporal modulation, the stochastic jitter in the propagation time of the signals is interpreted as an additive channel noise. In our system, we assume this delay jitter constitutes the sole source of uncertainty regarding the arrival times of the transmitted pulses. Physically, this delay is not necessarily calculated based on the peak concentration time; rather, it represents the triggering time—the moment when a sufficient number of neurotransmitter molecules successfully bind to the post-synaptic receptors to cross the activation threshold. Following the physical end-to-end models and the temporal modulation capacity bounds rigorously analyzed in \cite{aghababaiyan2018capacity}, we characterize this triggering temporal noise as an additive Gamma random variable \cite{karatzas2012brownian} with the following probability density function (PDF):
	\begin{equation}
		{f_{\delta}}(\delta) = \frac{{{\beta ^\alpha }}}{{\Gamma (\alpha )}}{\delta^{\alpha - 1}}{e^{ - \beta \delta}},
		\label{E14}
	\end{equation}
	where $\delta$ represents the channel noise (delay jitter), and $\alpha$ and $\beta$ are shape and rate parameters, respectively. 
	
	Consequently, the conditional PDF of the arrival time $Y = y$ given that the input signal was emitted at time $X = x$ is formulated based on the standard properties of the Gamma distribution as:
	\begin{equation}
		{f_{Y|X}}(y|x) = \left\{ {\begin{array}{*{20}{l}}
				{\frac{{{\beta ^\alpha }}}{{\Gamma (\alpha )}}{{(y - x)}^{\alpha - 1}}{e^{ - \beta (y - x)}},{\mkern 1mu} {\mkern 1mu} {\mkern 1mu} {\mkern 1mu} {\mkern 1mu} y > x,}\\
				{0, \,\,\,\,\,\,\,\,\,\,\,\,\,\,\,\,\,\,\,\,\,\,\,\,\,\,\,\,\,\,\,\,\,\,\,\,\,\,\,\,\,\,\,\,\,\,\,\,\,\,\,\,\,\,\,  y \le x.}
		\end{array}} \right.
		\label{E15}
	\end{equation}

	\section{Neuro-Spike Array-Based Communication Scheme}
	In this section, we propose a novel neuro-spike array-based communication scheme designed to bridge severed connections between damaged neurons using auxiliary nano-machines. Subsequently, we rigorously evaluate the performance of the proposed scheme by analyzing the probability of inter-symbol interference (ISI) occurrence and the achievable communication rate.
	
	\subsection{Proposed Array-Based Communication Scheme}
	Generally, in digital modulation schemes where the transmission alphabet comprises $l$ distinct symbols, the number of possible codewords of length $k$ is $l^k$. Let us consider the binary case ($l=2$), which has been widely utilized in the literature for modeling natural neuro-spike communications \cite{ref999, ref5new}. Hence, we assume the alphabet contains two symbols, i.e., \{`0', `1'\}, resulting in $2^k$ different codewords for a transmitted sequence of $k$ symbols.
	
\begin{figure}[t]
		\centering
		\begin{subfigure}[t]{0.64\textwidth}
			\centering
			\includegraphics[width=1\textwidth]{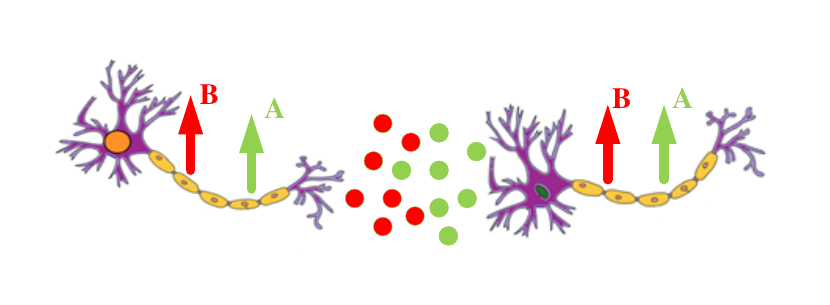}
			\caption{}
		\end{subfigure}%
		~
		\begin{subfigure}[t]{0.32\textwidth}
			\centering
			\includegraphics[width=1\textwidth]{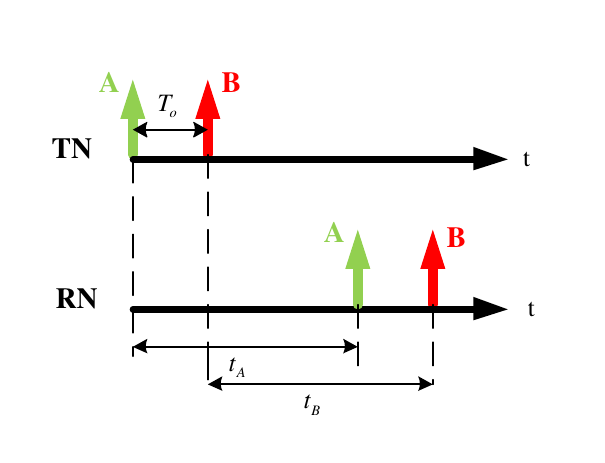}
			\caption{}
		\end{subfigure}%
		\caption{(a) The schematic of the proposed array-based neuro-spike communication scenario. (b) The time diagram of the binary scenario. The programmed intermission time between consecutive pulses is denoted by $T_o$. If the first pulse is emitted at time $t$, the second one is emitted at time $t+T_o$.}
		\label{12}
	\end{figure}

	Fig. \ref{12}(a) illustrates the schematic of the proposed scheme, where information is encoded into the sequential arrangement of emitted pulses. In this architecture, we consider two distinct types of molecular pulses. The transmitting nano-machines (TN) exploit different neurotransmitter molecules for release into the synaptic cleft. Specifically, we assume Glutamate and GABA as the signaling molecules, given their roles as the primary fast neurotransmitters in the mammalian cortex and their characteristically short reuptake periods.
	
	We denote an emitted pulse by $A$ when the internal circuit voltage $V_c(t)$ exceeds the upper threshold $V_{\text{th}1}$, and by $B$ when $V_c(t)$ drops below the lower threshold $V_{\text{th}2}$. Because these pulses consist of distinct neurotransmitter molecules, the receiving nano-machine (RN) can distinguish them unequivocally via specific chemical receptors, completely bypassing the need to measure exact amplitudes or precise arrival times. 
	
	Without loss of generality, we associate the transmission sequence $\{A-B\}$ with the input bit $x=0$, and the sequence $\{B-A\}$ with the input bit $x=1$. That is, to transmit a bit `1', the TN first emits pulse $A$ followed by pulse $B$. Conversely, to transmit a bit `0', it emits pulse $B$ followed by pulse $A$. The programmed intermission time between these two consecutive pulses is denoted by $T_o$. Hence, if the first pulse is released at time $t$, the second pulse is released at time $t+T_o$. Fig. \ref{12}(b) presents the time diagram of this binary scenario. 
	
	In addition, within the broader network scenario, the auxiliary nano-machines must also interconnect with healthy neurons. This interface can be achieved using standard on/off keying (OOK) modulation \cite{malak2013communication}. Under OOK, healthy neurons naturally transmit an action potential (spike) within a symbol duration to convey bit `1', and remain silent to convey bit `0'. Therefore, an RN can extract information from a healthy pre-synaptic neuron simply by scanning the symbol duration for any activity. Similarly, a TN can transmit information to a healthy post-synaptic neuron by artificially inducing a single spike within the symbol duration.
	
	To characterize the propagation delay of each molecular pulse traversing the synaptic cleft, we apply the Gamma distribution model established in (\ref{E14}). Thus, the propagation delay of any pulse to reach the RN obeys a Gamma PDF. In (\ref{E14}), $\Gamma (\alpha ) = \int\limits_0^\infty {{q^{\alpha - 1}}{e^{ - q}}dq}$, where the shape parameter $\alpha$ and the rate parameter $\beta$ are functions of the voltage thresholds and the diffusion coefficient $D$. The cumulative distribution function (CDF) of $f_{\delta}(\delta)$, denoted as $F_{\delta}(\delta)$, is obtained as:
	\begin{equation}
		{F_{\delta}}(\delta) = \frac{{\gamma (\alpha ,\beta \delta)}}{{\Gamma (\alpha )}},
		\label{E_CDF}
	\end{equation}
	where $\gamma (\alpha,\beta \delta) = \int\limits_0^{\beta \delta} {{q^{\alpha - 1}}{e^{ - q}}dq}$ is the lower incomplete Gamma function. In the following subsections, we mathematically analyze the binary array-based scheme in terms of the probability of ISI occurrence and the achievable communication rate.
	
	\subsection{The Probability of Successful Transmission}
	In this subsection, we derive the probability of a successful transmission for the binary array-based scheme, defined as the probability that a transmitted symbol is correctly decoded by the RN.
	
	\textbf{Theorem 1:}
	The probability that symbol $x=0$ is transmitted by the TN and correctly received by the RN is derived as:
	\begin{equation}
		{{\text{Pr}}}(x=0|y=0) =
		\frac{{\gamma (\alpha ,\beta {T_o})}}{{\Gamma (\alpha )}} + \frac{{\Gamma (\alpha ,\beta {T_o})}}{{\Gamma (\alpha )}}\int\limits_{{T_o}}^\infty {{f_{{\delta}}}({t})\frac{{\Gamma (\alpha ,\beta ({t} - {T_o}))}}{{\Gamma (\alpha )}}} d{t}.
		\label{E21}
	\end{equation}
	
	\textbf {Proof:} The proof is provided in Appendix A.
	
	\subsection{Inter-symbol Interference of Array-Based Communication Scheme}
	Consecutive symbols may interfere if the delay jitter causes their constituent pulses to arrive out of order at the RN, a phenomenon known as inter-symbol interference (ISI) which inherently causes transmission errors. In this subsection, we derive the probability of ISI for the proposed binary array-based scheme. 
	
	Because the propagation delays $t_A$ and $t_B$ follow the identical Gamma distribution, the probability of ISI occurrence is statistically equivalent for any transmitted sequence, regardless of whether pulse $A$ or $B$ is emitted first. Therefore, without loss of generality, we consider the transmission of the sequence `000' and evaluate the interference caused by adjacent symbols (both preceding and succeeding). We derive the probability of ISI occurrence, denoted as $P_{\text{ISI}}$, as follows.
	
	\begin{figure}[t]
		\centering
		\includegraphics[width=1\textwidth]{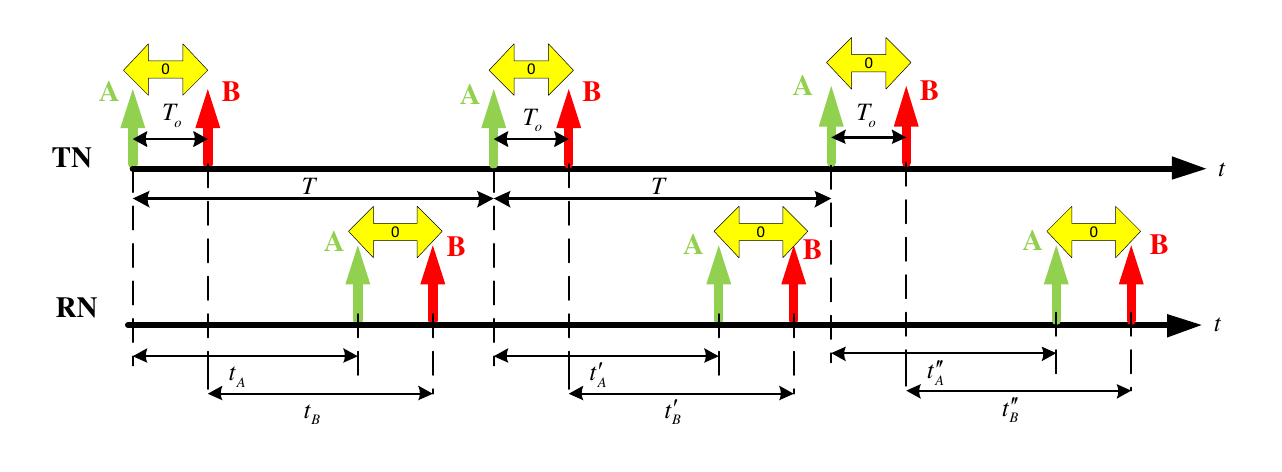}
		\caption{The relevant delay variables across three symbol durations required to derive the probability of ISI occurrence when the sequence `000' is transmitted.}
		\label{pic-7}
	\end{figure}
	
	Fig. \ref{pic-7} illustrates the relevant delay variables across three symbol durations utilized for deriving the ISI probability. In the first symbol, $t_A$ and $t_B$ denote the random propagation delays of pulses $A$ and $B$, respectively. The symbol duration is denoted by $T$. The maximum delay of pulses $A$ and $B$ can be specified by the random variable $\varphi$, where $\varphi = \max ({t_A},{t_B} + {T_o})$.
	
	Because $t_A$ and $t_B$ share the identical PDF given in (\ref{E14}), the PDF and CDF of the random variable $\varphi$ can be formulated using $f_{\delta}(\delta)$ and $F_{\delta}(\delta)$. Given that $t_A$ and $t_B+T_o$ are independent, the CDF of $\varphi$, i.e., $F_\varphi(t)$, is derived as:
	\begin{equation}
		{F_\varphi}(t) = {F_{{t_A}}}(t){F_{{T_o} + {t_B}}}(t),\,\,\,\,\,\,\,\,\,\,\,\,\,\,\,\,\,\, t \ge {T_o},
		\label{E22}
	\end{equation}
	where ${F_{{T_o} + {t_B}}}(t)$ is obtained by time-shifting ${F_{{t_B}}}(t)$, such that ${F_{{T_o} + {t_B}}}(t) = {F_{{t_B}}}(t - {T_o})$. Since $t_A$ and $t_B$ are identically distributed, i.e., ${F_{{t_A}}}(t)={F_{{t_B}}}(t)={F_{{\delta}}}(t)$, the expression in (\ref{E23}) can be expanded as:
	\begin{equation}
		{F_\varphi}(t) = {F_{{t_A}}}(t){F_{{t_B}}}(t - {T_o}) = \frac{1}{{{\Gamma ^2}(\alpha )}}\gamma (\alpha ,\beta t)\gamma (\alpha ,\beta (t - {T_o})).
		\label{E23}
	\end{equation}
	By differentiating the expression in (\ref{E23}), the PDF of $\varphi$, i.e., $f_\varphi(t)$, is obtained as:
	\begin{equation}
		{f_\varphi}(t) = f_{\delta}(t)F_{\delta}(t - {T_o}) + F_{\delta}(t)f_{\delta}(t - {T_o}),\,\,\,\,\,\,\,\,\,\,\,\,\,\,\,\,\,\,t \ge {T_o}.
		\label{E24}
	\end{equation}
	
	For the second symbol duration, the delays of pulses $A$ and $B$ are denoted as ${t'_A}$ and ${t'_B}$, respectively. The minimum and maximum delays of the pulses in this symbol are characterized by the random variables $\varphi '$ and $\chi$. Specifically, we define $\chi = \max ({t'_A} + T,{t'_B} + {T_o} + T)$ and $\varphi ' = \min ({t'_A} + T,{t'_B} + {T_o} + T)$. Since the second bit is transmitted after one symbol duration $T$, the CDF of $\chi$, denoted by $F_\chi(t)$, is given by:
	\begin{equation}
		{F_\chi}(t) = {F_{{t_A}}}(t-T){F_{{T_o} + {t_B}}}(t-T), \,\,\,\,\,\,\,\,\,\,\,\,\,\,\,\, t \ge {T_o+T}.
		\label{E25}
	\end{equation}
	This expression can be simplified to:
	\begin{equation}
		{F_\chi}(t) = {F_{{\delta}}}(t-T){F_{{\delta}}}(t - {T_o}-T) = \frac{1}{{{\Gamma ^2}(\alpha )}}\gamma (\alpha ,\beta (t-T))\gamma \left( \alpha ,\beta (t - {T_o}-T)\right).
		\label{E26}
	\end{equation}
	Similarly, the CDF of $\varphi ' = \min ({t'_A} + T,{t'_B} + {T_o} + T)$, i.e., ${F_{\varphi '}}(t)$, is derived as:
	\begin{equation}
		{F_{\varphi '}}(t) = F_{\delta}(t - T) + F_{\delta}(t - {T_o} - T)- F_{\delta}(t - T)F_{\delta}(t - {T_o} - T), \,\,\,\,\,\,\,\,\,\,\,\,\,\,\,\,t \ge T.
		\label{E27}
	\end{equation}
	Which can be rewritten in terms of the incomplete Gamma function:
	$${F_{\varphi '}}(t) = \frac{1}{{\Gamma (\alpha )}}\left[ {\gamma (\alpha ,\beta (t - T)) + \gamma (\alpha ,\beta (t - {T_o} - T))} \right]$$
	\begin{equation}
		\ \ \ \ \ - \frac{1}{{{\Gamma ^2}(\alpha )}}\left[ {\gamma (\alpha ,\beta (t - T))\gamma (\alpha ,\beta (t - {T_o} - T))} \right], \,\,\,\,\,\,t \ge T.
		\label{E28}
	\end{equation}
	By differentiating ${F_{\chi}}(t)$, the PDF of ${\chi}$, i.e., ${f_{\chi}}(t)$, is obtained as:
	\begin{equation}
		{f_\chi}(t) = f_{\delta}(t - 2T)\frac{{\Gamma (\alpha ,t - {T_o} - 2T)}}{{\Gamma (\alpha )}}+ f_{\delta}(t - {T_o} - 2T)\frac{{\Gamma (\alpha ,t - 2T)}}{{\Gamma (\alpha )}}, \,\,\,\,\,\,\,\,\,\,\,\,\,\,\,\,t \ge 2T.
		\label{E29}
	\end{equation}
	Likewise, differentiating ${F_{\varphi '}}(t)$ yields the PDF of ${\varphi '}$:
	\begin{equation}
		{f_{\varphi '}}(t) = f_{\delta}(t - T)\frac{{\Gamma (\alpha ,t - {T_o} - T)}}{{\Gamma (\alpha )}}+ f_{\delta}(t - {T_o} - T)\frac{{\Gamma (\alpha ,t - T)}}{{\Gamma (\alpha )}}, \,\,\,\,\,\,\,\,\,\,\,\,\,\,\,\,t \ge T.
		\label{E30}
	\end{equation}
	
	Finally, we define $\chi '$ as the random variable representing the minimum delay of pulses $A$ and $B$ in the third symbol duration (for $x=0$), such that $\chi ' = \min ({t''_A} + 2T,{t''_B} + {T_o} + 2T)$. Because the third bit is transmitted after a delay of $2T$, the CDF of $\chi '$, i.e., ${F_{\chi '}} (t)$, is derived by substituting $2T$ for $T$ in ${F_{\varphi '}} (t)$:
	$${F_{\chi '}}(t) = \frac{1}{{\Gamma (\alpha )}}\left[ {\gamma (\alpha ,\beta (t - 2T)) + \gamma (\alpha ,\beta (t - {T_o} -2 T))} \right]$$
	\begin{equation}
		- \frac{1}{{{\Gamma ^2}(\alpha )}}\left[ {\gamma (\alpha ,\beta (t - 2T))\gamma (\alpha ,\beta (t - {T_o} - 2T))} \right],\,\,\,\,\, t \ge 2T.
		\label{E31}
	\end{equation}
	
	\textbf{Theorem 2:} The probability of ISI occurrence in the binary array-based scheme is analytically derived as:
	$${P_{\text{ISI}}} = 1- \left( {\int\limits_{ - \infty }^{ - T} {\int\limits_m^\infty {f_{\delta}(v){f_s}(t - v)dvdt} } } \right) $$
	$$\times \left[ {1 - \frac{{\gamma \left( {\alpha ,\beta {T_o}} \right)}}{{\Gamma (\alpha )}} - \frac{{\Gamma (\alpha ,\beta {T_o})}}{{\Gamma (\alpha )}}\int\limits_{{T_o}}^\infty {f_{\delta}({t})\frac{{\Gamma \left( {\alpha ,\beta ({t} - {T_o})} \right)}}{{\Gamma (\alpha )}}} d{t}} \right] $$
	$$ - \left( {\int\limits_{ - \infty }^{ - {T_o} - T} {\int\limits_m^\infty {f_{\delta}(v){f_s}(t - v)dvdt} } } \right)$$
	$$ \times \left[ {\frac{{\gamma (\alpha ,\beta {T_o})}}{{\Gamma (\alpha )}} + \frac{{\Gamma (\alpha ,\beta {T_o})}}{{\Gamma (\alpha )}}\int\limits_{{T_o}}^\infty {f_{\delta}({t})\frac{{\Gamma (\alpha ,\beta ({t} - {T_o}))}}{{\Gamma (\alpha )}}d{t}} } \right]$$
	\begin{equation}
		\times \left( {1 - \int\limits_{ - \infty }^0 {\int\limits_n^\infty {{f_{\varphi'}}(v){f_r}(t - v)dvdt} } } \right),
		\label{E48}
	\end{equation}
	where $f_{ \varphi'}(t)$, $f_s(t)$, and $f_r(t)$ are given in (\ref{E30}), (\ref{E36}), and (\ref{E43}), respectively.
	
	\textbf {Proof:} The detailed proof is provided in Appendix B.

	\subsection{Error Probability and Achievable Rate of Binary Array-Based Communication Scheme}
	In this subsection, we evaluate the overall performance of the binary array-based communication scheme by deriving its error probability and the resulting achievable communication rate per transmission.
	
	\textbf{Theorem 3:}
	The bit error probability of the proposed binary array-based communication scheme, $P_e$, is obtained as:
	\begin{equation}
		P_e = 1 - P_{\text{NoISI}|(x=0|y=0)}{{\text{Pr}}}(x=0|y=0)= 1 - I({T_o} + T)\left( {1 - J(T)} \right){{\text{Pr}}}(x=0|y=0),
		\label{E58}
	\end{equation}
	where $I(T) = \int\limits_{ - \infty }^{ - T} {{f_{ {t_A}+s}}(t)dt}$ and $J(T) = \int\limits_{ - \infty }^{ - T} {{f_{{t_A} + r}}(t)dt}$.
	
	\textbf {Proof:} The proof is provided in Appendix C.
	
	By substituting the explicit formulations for $I({T_o} + T)$, ${J(T)}$, and ${{\text{Pr}}}(x=0|y=0)$ into (\ref{E58}), $P_e$ is expanded as:
	$$P_e =1- \left[ {\int\limits_{ - \infty }^{ - {T_o} - T} {\int\limits_m^\infty {f_{\delta}(v){f_s}(t - v)dvdt} } } \right]\times \left[ {1 - \int\limits_{ - \infty }^{ - T} {\int\limits_h^{ + \infty } {f_{\delta}(v){f_r}(t - v)dvdt} } } \right] $$
	\begin{equation}
		\times \left[ {\frac{{\gamma (\alpha ,\beta {T_o})}}{{\Gamma (\alpha )}} + \frac{{\Gamma (\alpha ,\beta {T_o})}}{{\Gamma (\alpha )}}\int\limits_h^{ + \infty } {f_{\delta}({t})\frac{{\Gamma \left( {\alpha ,\beta ({t} - {T_o})} \right)}}{{\Gamma (\alpha )}}} d{t}} \right],
		\label{E60}
	\end{equation}
	where $f_s(t)$ and $f_r(t)$ are given in (\ref{E36}) and (\ref{E43}), respectively. The integration bounds are defined as $h = \max \left( {0,t + {T_o}} \right)$ and $m = \max (0,t + 2T).$
	
	In the proposed array-based scheme, the TN transmits binary symbols (`0' or `1'), and the RN successfully decodes the information bit with probability $1-P_e$, where $P_e$ represents the bit error probability. Consequently, the achievable rate of the proposed scheme in bits per transmission, denoted by $C_\text{Array}$, is given by $C_\text{Array}=1-H(P_e)$, where $H(P) = - P{\log _2}(P) - (1 - P){\log _2}(1 - P)$ is the binary entropy function. Substituting the expression for $P_e$ into this entropy equation, the achievable rate is formulated as:
	\begin{equation}
		C_\text{Array} =1-H\left( I({T_o} + T)\left( {1 - J(T)} \right){{\text{Pr}}}(x=0|y=0) \right).
		\label{E59}
	\end{equation}
	This closed-form expression characterizes the achievable capacity of the proposed architecture in bits per transmission. To evaluate the throughput in the time domain, the communication rate in bits per second (bps) is calculated simply as ${C_\text{Array}}/{T}$.
	
	\section{Numerical Results}
	In this section, we present numerical results to rigorously evaluate the performance of the proposed neuro-spike array-based communication scheme. Specifically, we investigate the theoretical expressions derived for the probability of ISI occurrence (\ref{E48}) and the achievable communication rate (\ref{E59}). Furthermore, we analyze the impact of the symbol duration ($T$) and the programmed intermission time ($T_o$) on system performance. In our analysis, the symbol duration is varied within the interval of $1$ to $6 \, \textrm{ms}$. Since the intermission time must naturally be shorter than the symbol duration, $T_o$ is strictly bounded by $T$. All numerical computations and simulations in this section are executed using MATLAB.
	
	\begin{table}[t]
		\centering
		\caption{Simulation Parameters}
		\label{tab-1}
		\begin{centering}
			\scalebox{0.87}{
				\begin{tabular}{ |c|c|c|  }
					\hline Parameter & Symbol & Value\\
					\hline $\text{I}\&\text{F}$ logic circuit time constant & $\tau_{m}$ & $10 \,\textrm{ms}$\\
					\hline Upper threshold voltage (Pulse A) & $V_{\text{th}1}$ & $-55 \,\textrm{mv}$\\
					\hline Lower threshold voltage (Pulse B) & $V_{\text{th}2}$ & $-85 \,\textrm{mv}$\\
					\hline Initial voltage of capacitance & $V_c(0)$ & $-70 \,\textrm{mv}$\\
					\hline Diffusion coefficient & $D$ & $0.1, 0.3, 0.5 \, \mu \textrm{m}^{2}/ \textrm{ms}$\\
					\hline Symbol duration & $T$ & Evaluated up to $6 \,\textrm{ms}$\\
					\hline Programmed intermission time & $T_o$ & Varied up to $T$\\
					\hline
			\end{tabular}}
		\end{centering}
	\end{table}
	
\begin{figure}[t]
		\centering
		\includegraphics[width=0.75\textwidth]{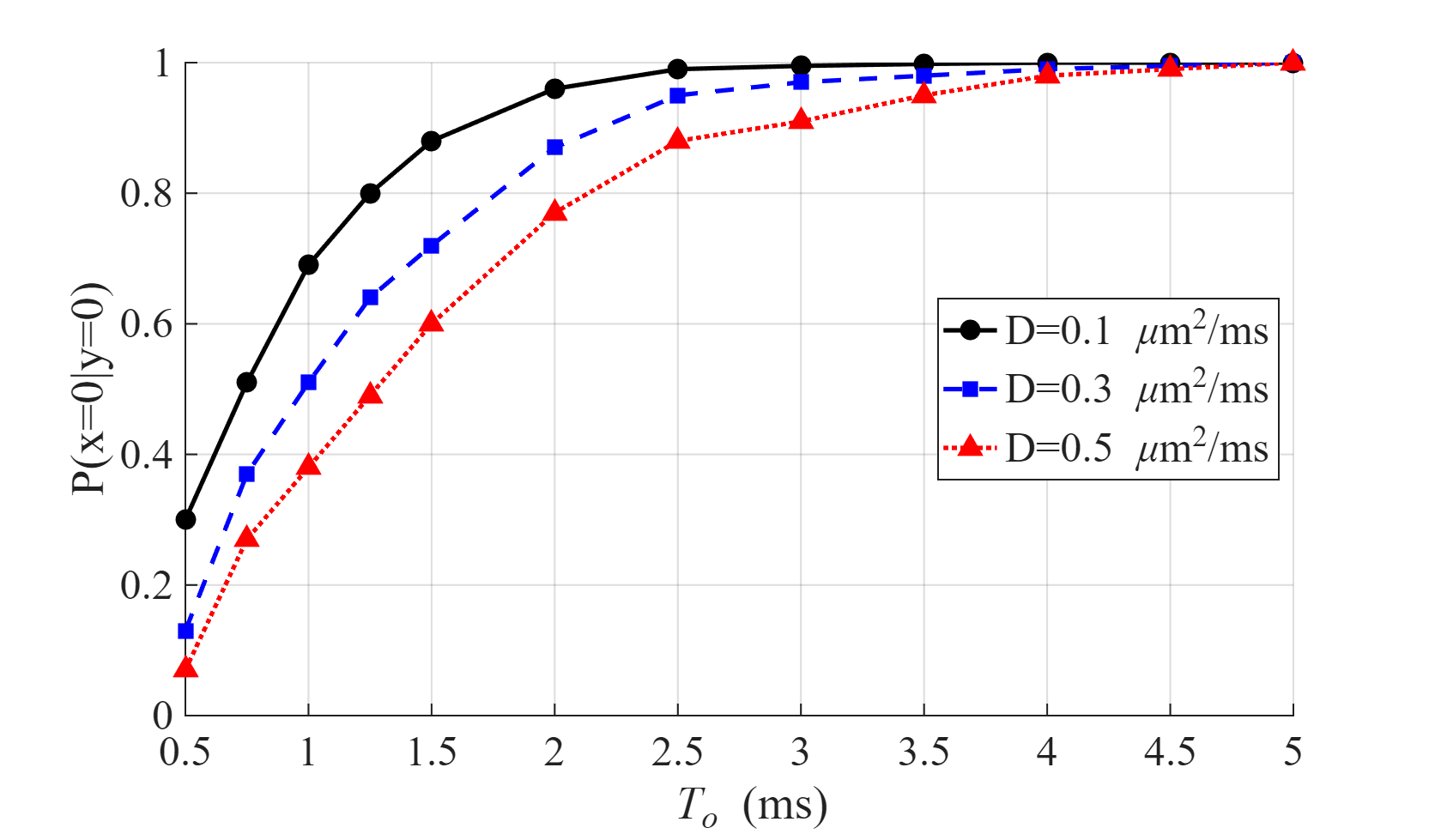}
		\caption{The successful transmission probability versus intermission time, i.e., $T_o$, when bit $x=0$ is transmitted for different values of the diffusion coefficient $D$.}
		\label{131}
	\end{figure}

	Fig. \ref{131} illustrates the successful transmission probability when bit $x=0$ is transmitted, i.e., ${{\text{Pr}}}(x=0|y=0)$, as a function of the intermission time $T_o$ for different values of the diffusion coefficient $D$. The remaining system parameters are configured according to Table \ref{tab-1}. It is evident that the probability of correct reception improves monotonically as $T_o$ increases. This occurs because a larger temporal gap between the two molecular pulses reduces the likelihood of them arriving out of order. Moreover, the curves indicate that to achieve a specific target probability of successful reception (e.g., $P \approx 0.8$), environments with higher values of $D$ (which correspond to faster diffusion dynamics) inherently require a larger intermission time $T_o$. It is also worth noting that based on the analytical derivation in Theorem 1, this successful reception probability solely depends on $T_o$ and is independent of the overall symbol duration $T$.
	
\begin{figure}[t]
		\centering
		\includegraphics[width=0.75\textwidth]{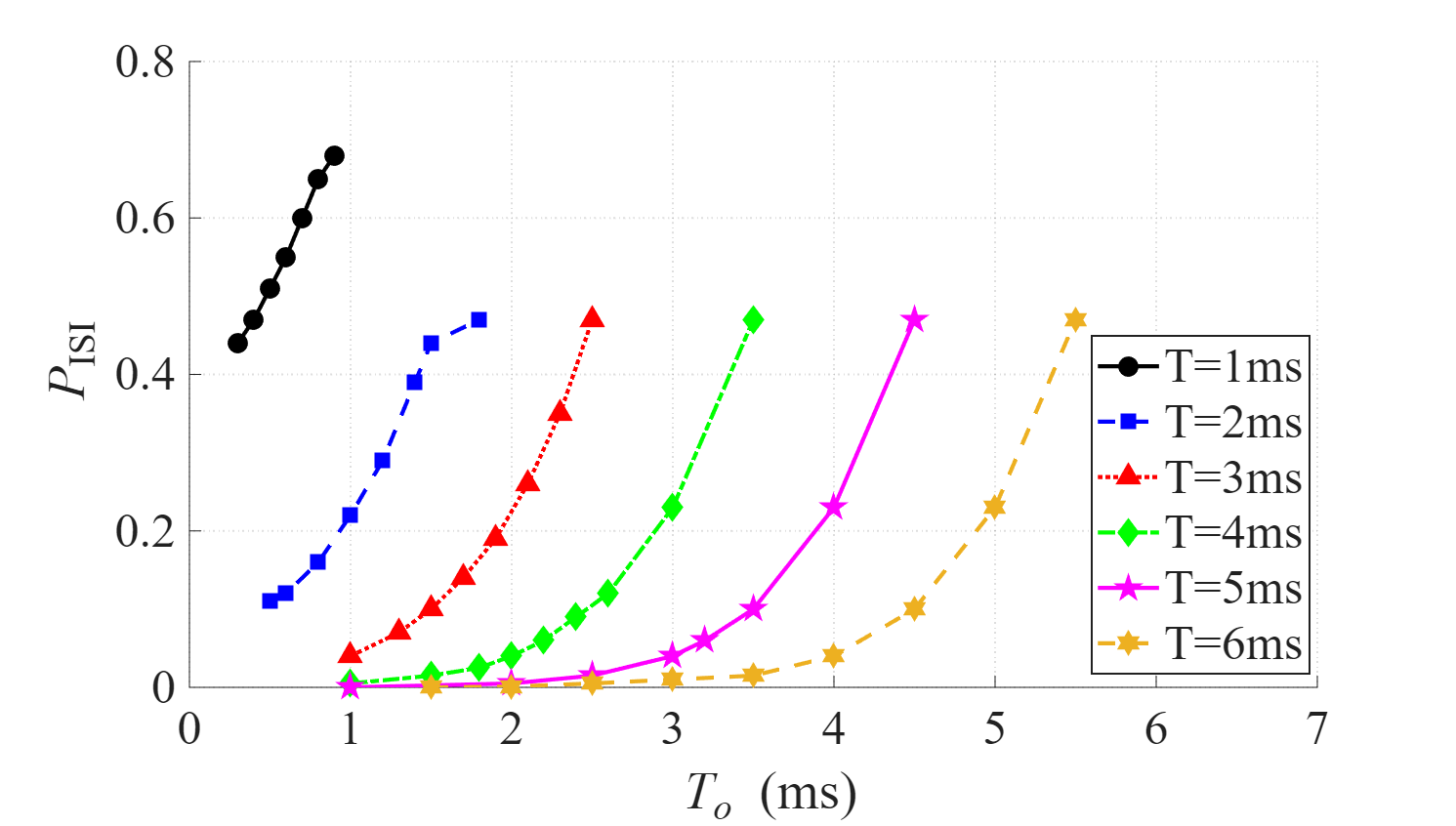}
		\caption{The probability of ISI occurrence with respect to $T_o$ when $D=0.3 \, \mu \textrm{m}^{2}/ \textrm{ms}$.}
		\label{132}
	\end{figure}

	Fig. \ref{132} depicts the probability of ISI occurrence ($P_{\text{ISI}}$) with respect to $T_o$ for varying symbol durations $T$, assuming a fixed diffusion coefficient of $D=0.3 \, \mu \textrm{m}^{2}/ \textrm{ms}$. As the symbol duration expands, the occurrence of ISI decreases for any given intermission time. This improvement is intuitive, as a longer symbol duration ensures that the molecular pulses of the current symbol are highly likely to have arrived and cleared the channel before the subsequent symbol duration commences. This can be explicitly observed by fixing the intermission time on the x-axis (e.g., $T_o = 2 \, \text{ms}$) and comparing the curves vertically; the probability of ISI drops significantly as $T$ increases from $3 \, \text{ms}$ to $6 \, \text{ms}$.
	
	\begin{figure}[t]
		\centering
		\includegraphics[width=0.75\textwidth]{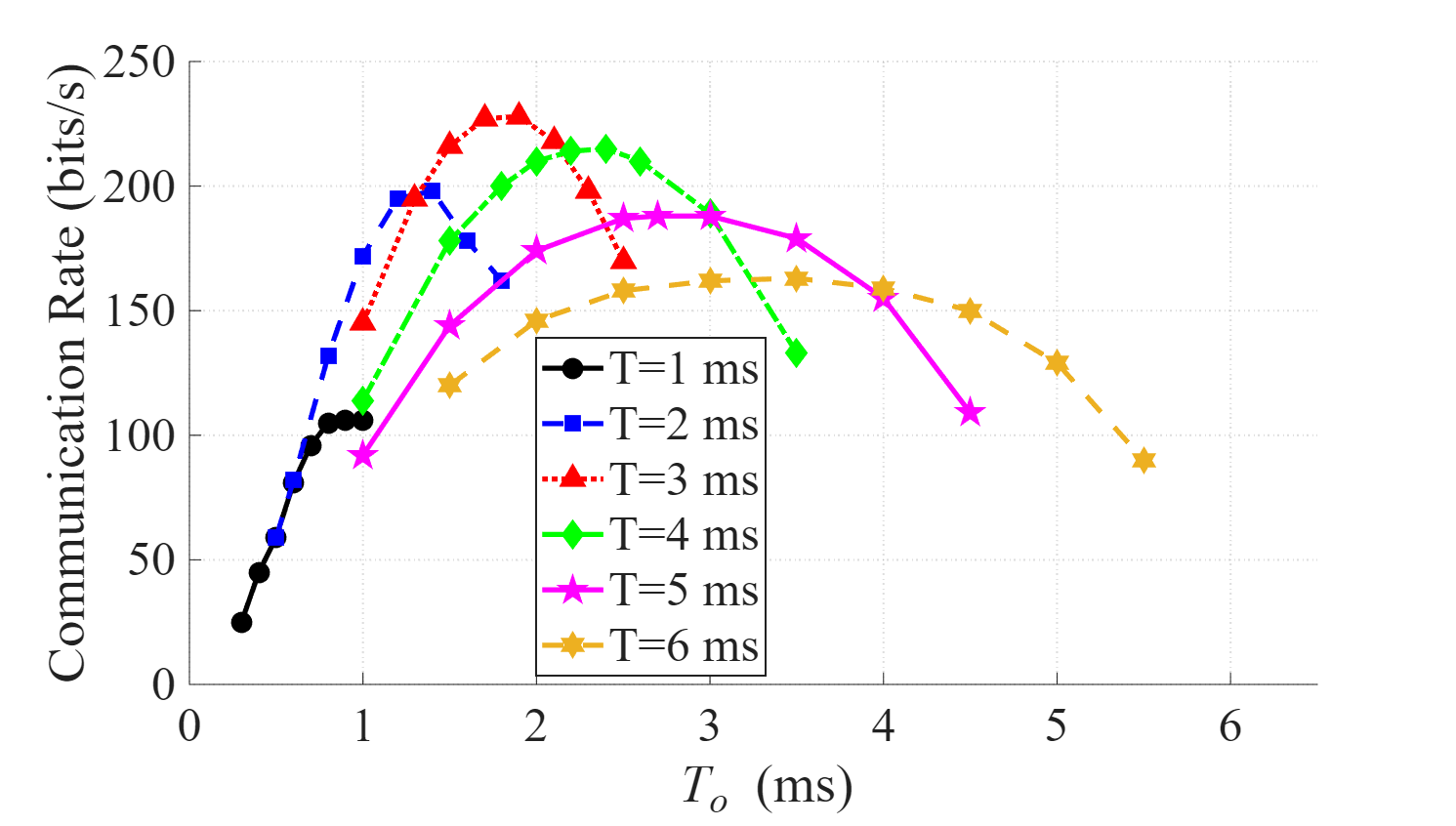}
		\caption{The achievable communication rate versus $T_o$ for different values of $T$, assuming $D=0.3 \, \mu \textrm{m}^{2}/ \textrm{ms}$.}
		\label{133}
	\end{figure}
	
	Conversely, it can be observed that for a fixed symbol duration, $P_{\text{ISI}}$ increases as $T_o$ grows. This is because enlarging $T_o$ shrinks the temporal guard band (the difference between $T$ and $T_o$), causing the trailing pulse of the current symbol to increasingly interfere with the leading pulse of the next symbol.
	
	Fig. \ref{133} demonstrates the achievable communication rate of the proposed scheme, $C_\text{Array}/T$, in bits per second (bps) for $D=0.3 \, \mu \textrm{m}^{2}/ \textrm{ms}$. The throughput $C_\text{Array}/T$ is governed by a fundamental trade-off between the bit error probability ($P_e$) and the symbol duration ($T$), since $C_\text{Array}=1-H(P_e)$. Initially, reducing $T$ enhances the overall communication rate because $T$ resides in the denominator. Physically, reducing $T$ means the transmitter emits symbols at a higher frequency. However, if $T$ drops below a critical threshold, the system suffers from ``temporal crowding''—a physical phenomenon where neurotransmitter molecules from consecutive pulses overlap and mix within the synaptic cleft due to insufficient clearance time. This physical overlap severely confuses the receiver's structural decoding, causing $P_{\text{ISI}}$ to surge, which heavily penalizes $P_e$ and consequently collapses the effective communication rate. For instance, in the scenario where $D=0.3 \, \mu \textrm{m}^{2}/ \textrm{ms}$, the absolute maximum communication rate is achieved at $T=3\, \textrm{ms}$ with an intermission time of $T_o=1.8 \,\textrm{ms}$.
	
	Furthermore, examining a specific symbol duration reveals that increasing $T_o$ initially reduces $P_e$, as a wider intermission provides a clearer decision boundary for the RN to distinguish between the arrangements of bit `0' and bit `1'. Nevertheless, as $T_o$ approaches the limits of $T$, the rate drops sharply due to the escalating $P_{\text{ISI}}$ (as corroborated by Fig. \ref{132}). Therefore, for every chosen symbol duration $T$, there exists a distinct optimal value of $T_o$ that perfectly balances internal symbol clarity against external inter-symbol interference, thereby maximizing the communication rate.
	
	\begin{figure}[t]
		\centering
		\includegraphics[width=0.75\textwidth]{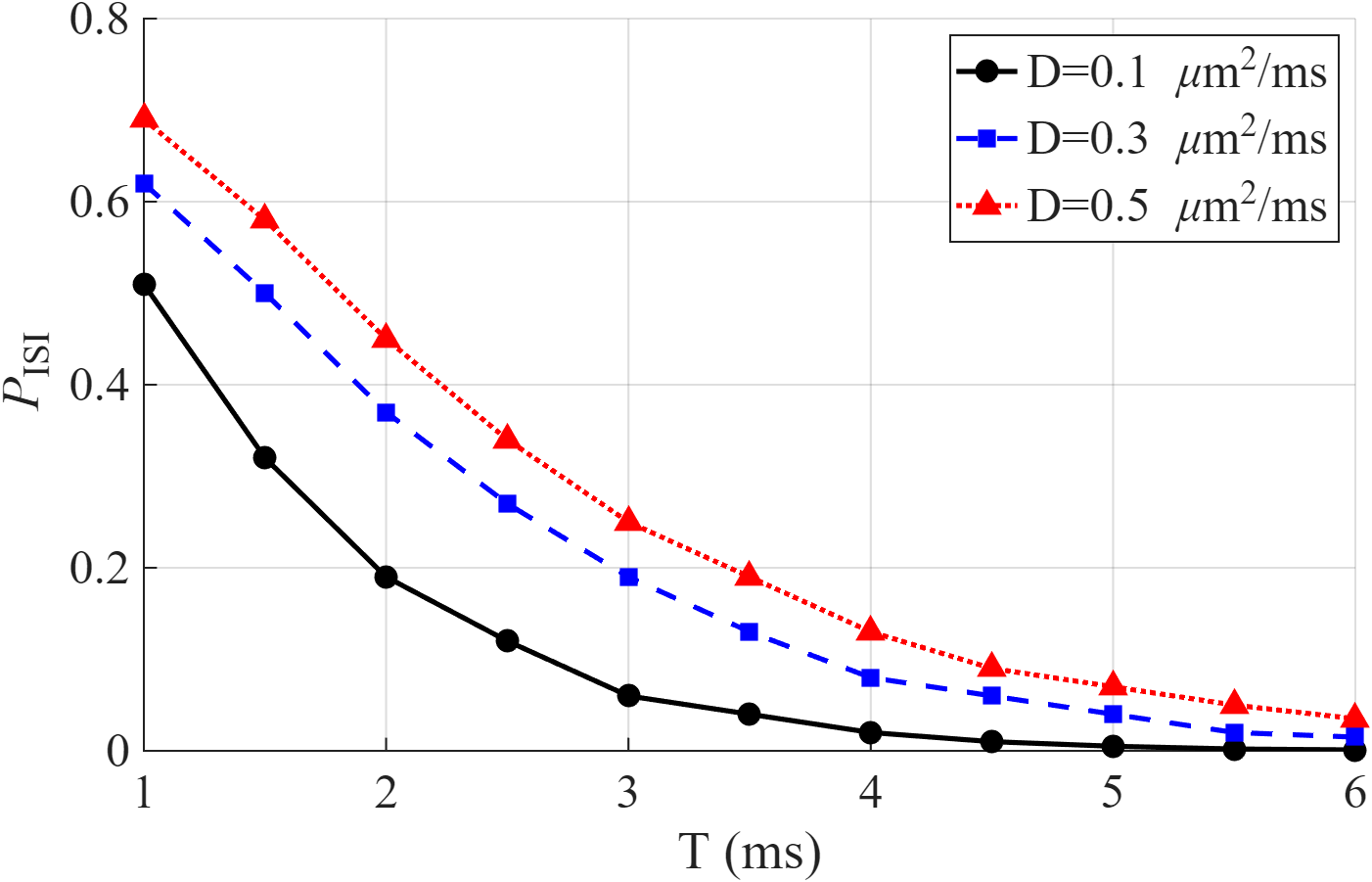}
		\caption{$P_{\text{ISI}}$ of the proposed scheme versus values of $T$ for $D=0.1, 0.3$, and $0.5 \, \mu \textrm{m}^{2}/ \textrm{ms}$.}
		\label{14}
	\end{figure}
	
	\begin{figure}[t]
		\centering
		\includegraphics[width=0.75\textwidth]{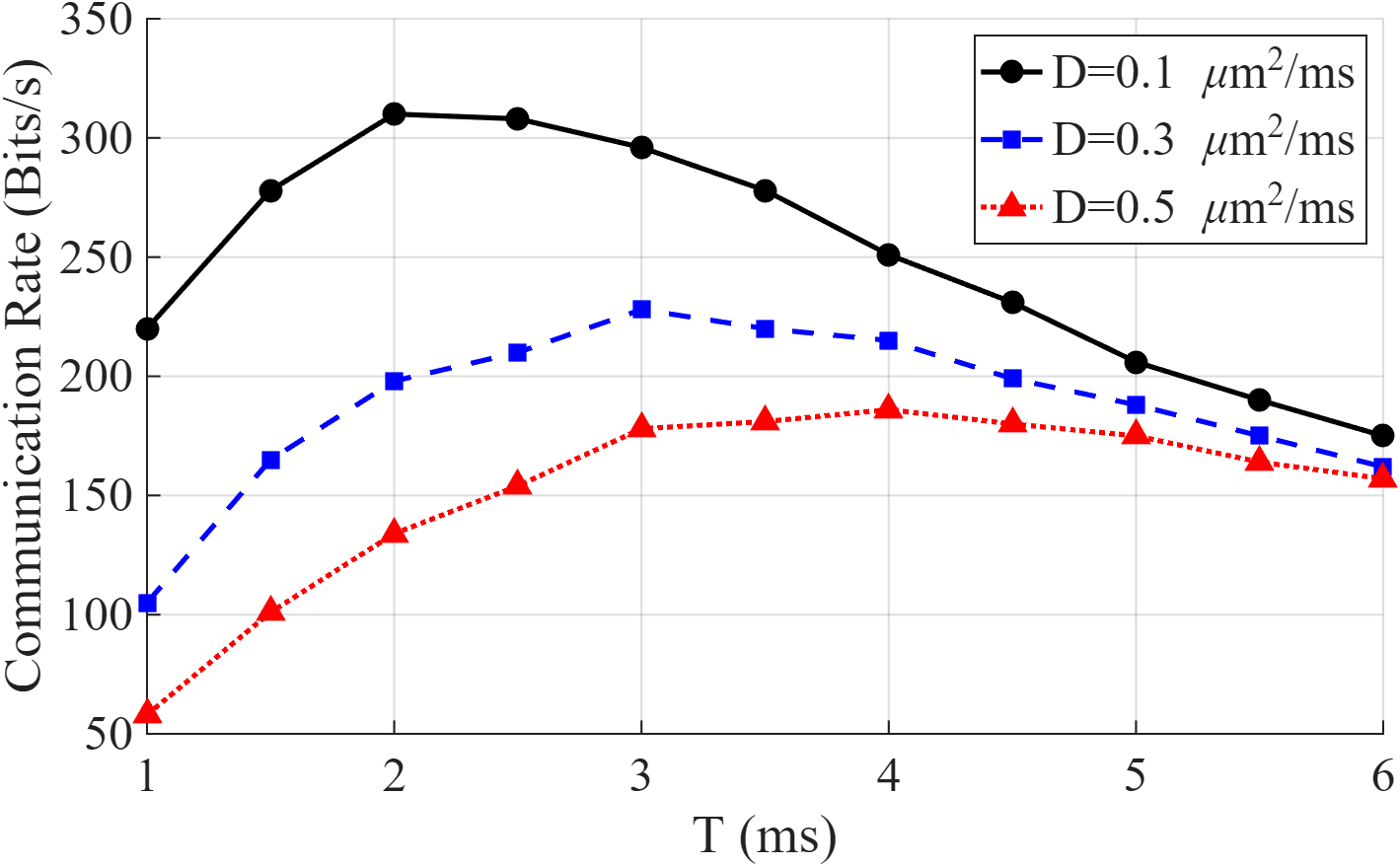}
		\caption{The communication rate of the proposed scheme versus values of $T$ for $D=0.1, 0.3$, and $0.5 \, \mu \textrm{m}^{2}/ \textrm{ms}$.}
		\label{144}
	\end{figure}
	
	Fig. \ref{14} plots $P_{\text{ISI}}$ versus $T$ across various diffusion coefficients. It is shown that $P_{\text{ISI}}$ is highly sensitive to $D$, increasing significantly under diffusion conditions that exacerbate the spatial spread and latency of neurotransmitters across the synaptic cleft. For example, observing the vertical cross-section at $T = 2 \, \text{ms}$, an environment with sluggish diffusion dynamics ($D = 0.1$) exhibits a lower ISI probability of approximately $0.2$. In contrast, under rapid diffusion conditions ($D = 0.5$) for the same symbol duration, the ISI probability increases to nearly $0.45$. This variation demonstrates the high sensitivity of the interference to the diffusion coefficient. Expectedly, $P_{\text{ISI}}$ declines as $T$ is extended. We can deduce that channels with more severe diffusion dynamics necessitate larger symbol durations to suppress ISI effectively.
	
	Fig. \ref{144} plots the optimized communication rate of the proposed scheme versus $T$ for different values of $D$. Unfavorable diffusion latencies directly throttle the maximum achievable rate. The curves clearly indicate the presence of an optimal $T$ for each value of $D$ that maximizes the throughput. We conclude that as the diffusion conditions worsen, the system must adapt by extending the symbol duration to maintain optimal communication efficiency.
	
\begin{figure}[htbp!] 
		\centering
		\includegraphics[width=0.75\textwidth]{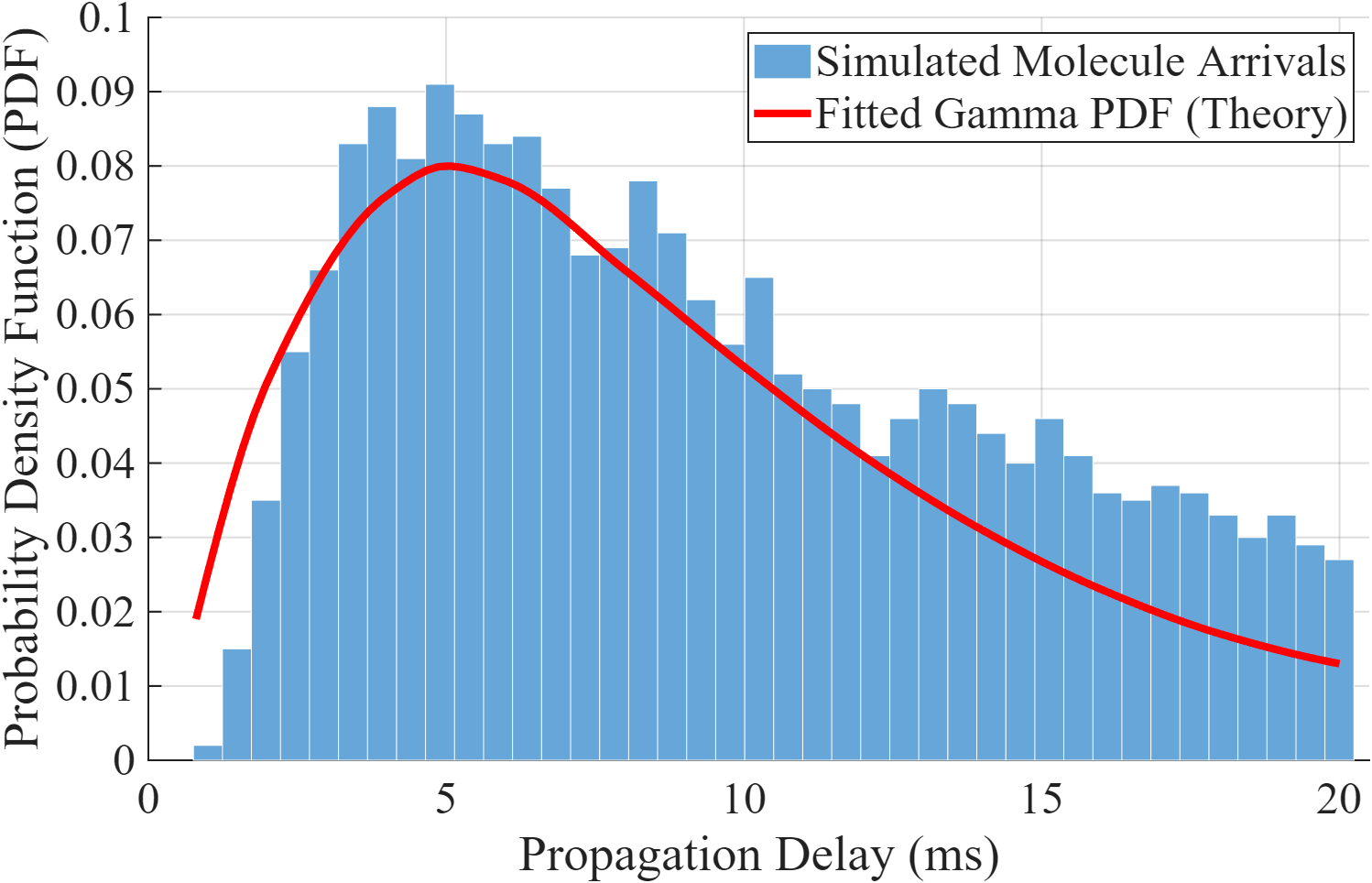} 
		\caption{3D Particle-based Brownian motion simulation validating the theoretical Gamma distribution of the propagation delay ($D=0.1 \, \mu \textrm{m}^{2}/ \textrm{ms}$).}
		\label{fig:val} 
	\end{figure}

	To validate the physical foundation of our theoretical framework—which fundamentally assumes that the propagation delay (temporal jitter) follows a Gamma distribution—we implemented a comprehensive 3D particle-based Brownian motion simulator. Fig. \ref{fig:val} demonstrates the recorded physical arrival times of 10,000 neurotransmitter molecules released into the synaptic cleft for an environment with $D = 0.1 \, \mu\text{m}^2/\text{ms}$. As illustrated, the empirical histogram of the simulated particle arrivals strongly aligns with the theoretical Gamma Probability Density Function (PDF) curve utilized throughout our analysis. Because the particle-based simulation validates this underlying physical delay model, it inherently corroborates the accuracy of the subsequent analytical bounds derived in Eqs. (15) and (18).
	
\begin{figure}[htbp!]
		\centering
		
		\begin{subfigure}[b]{1\textwidth} 
			\centering
			\includegraphics[width=0.55\textwidth]{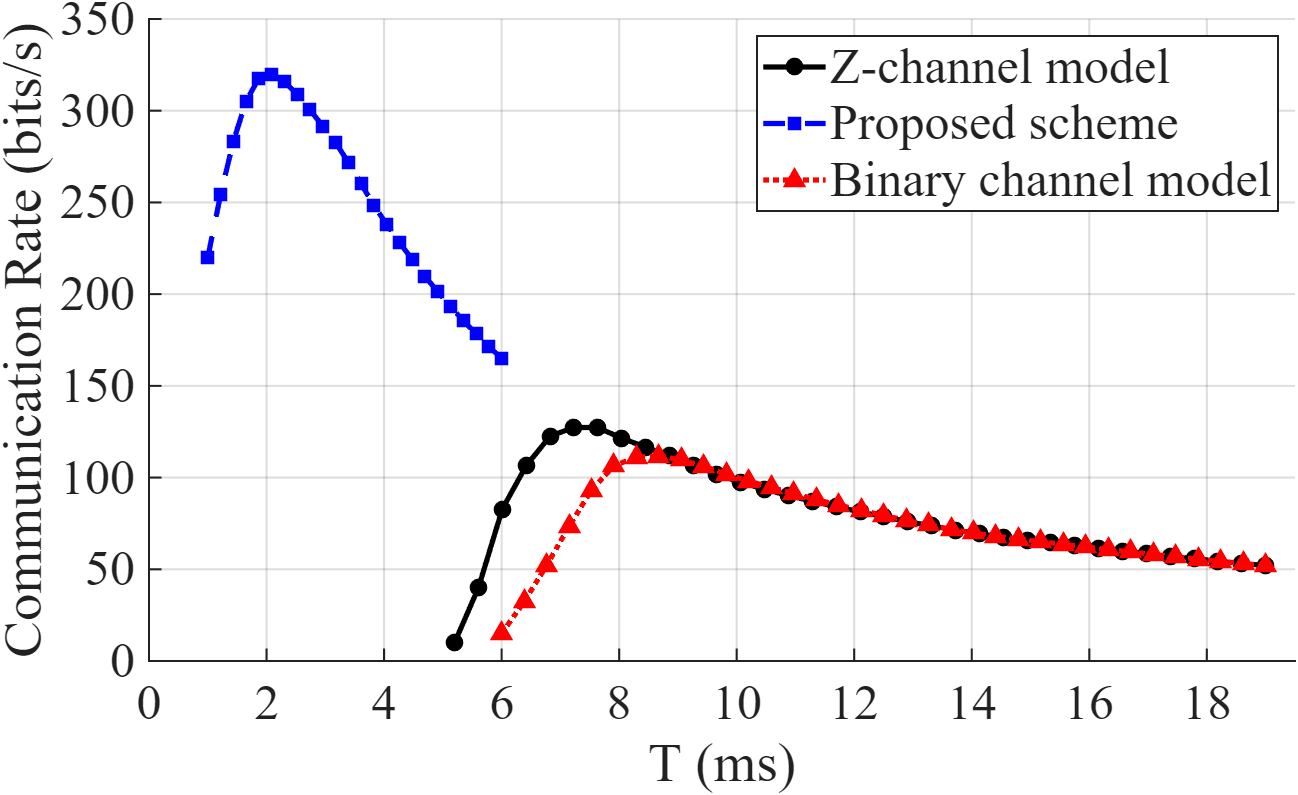} 
			\caption{}
		\end{subfigure}
		
		\vspace{0.5em} 
		
		\begin{subfigure}[b]{1\textwidth}
			\centering
			\includegraphics[width=0.55\textwidth]{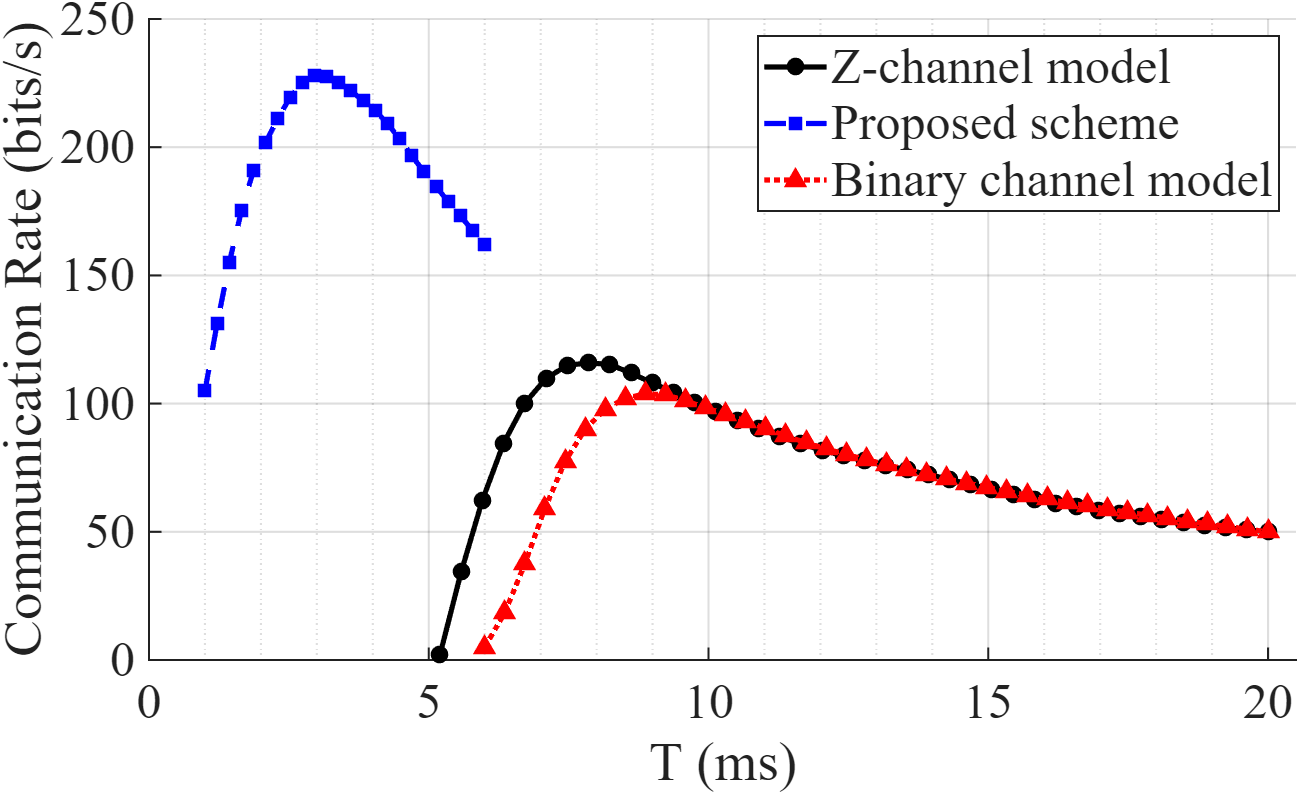}
			\caption{}
		\end{subfigure}
		
		\vspace{0.5em}
		
		\begin{subfigure}[b]{1\textwidth}
			\centering
			\includegraphics[width=0.55\textwidth]{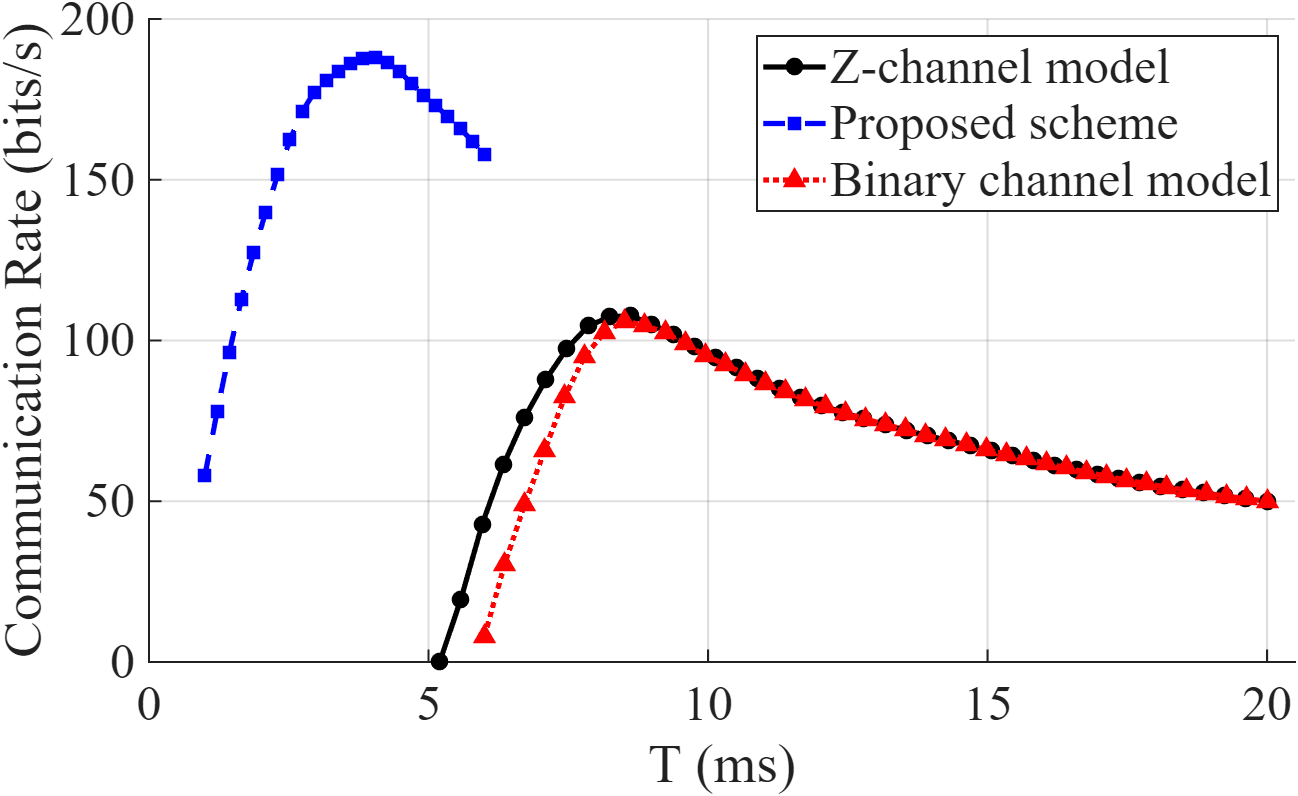}
			\caption{}
		\end{subfigure}
		
		\caption{Comparative analysis of the optimal communication rate operational ranges for the proposed array-based scheme and the baseline OOK models for (a) $D=0.1$, (b) $D=0.3$, and (c) $D=0.5 \, \mu \textrm{m}^{2}/ \textrm{ms}$.}
		\label{fig:15} 
	\end{figure}

	Finally, Fig. \ref{fig:15} benchmarks the communication rate of the proposed array-based scheme against the standard on-off keying (OOK) modulation operating over conventional symbol-synchronized Z-channel and binary channel models. The mathematical formulations for the capacities of these baseline models are detailed in Appendix D. It should be noted that the x-axis in Fig. \ref{fig:15} universally represents the exact same metric: the symbol duration $T$. The curves are plotted across different ranges of $T$ to highlight the optimal operational region of each scheme. Because our proposed array-based scheme efficiently suppresses ISI, it achieves its peak capacity at very short symbol durations ($T < 6$ ms). In contrast, the OOK baseline models suffer from severe ISI at such high transmission speeds and strictly require much longer symbol durations to clear the channel and achieve meaningful communication rates.
	
	To provide a deeper comparative insight into this performance disparity, it is crucial to analyze the fundamental limitations of the baseline schemes. The conventional OOK models are highly vulnerable to the heavy-tail property of the Gamma-distributed propagation delay. In the standard binary channel model, a logic `1' (pulse emission) followed by a logic `0' (silence) frequently results in a false positive, as delayed residual molecules from the first time slot spill over into the second. Conversely, our array-based scheme utilizes two chemically distinct neurotransmitters (i.e., Glutamate and GABA). Because the RN decodes symbols based on structural receptor binding rather than mere concentration thresholds, the trailing tail of a Glutamate pulse cannot falsely trigger a GABA receptor, and vice versa. This biochemical orthogonality inherently suppresses the devastating effects of ISI, allowing the proposed scheme to pack symbols much closer together in time.
	
	Furthermore, a comparative examination of the sensitivity to the diffusion coefficient reveals a pronounced resilience in the proposed architecture. When the diffusion environment worsens, the temporal spread of molecules increases drastically. As observed for the baseline models, they are forced to significantly increase their optimal symbol durations to maintain channel reliability and wait for the channel to clear. In stark contrast, the array-based encoding provides robust, high-speed neural stimulation that is far less susceptible to harsh channel variations than traditional amplitude-modulated benchmarks.

	\section{Discussion and Practical Limitations}
	While the proposed array-based molecular pulse encoding effectively circumvents stringent synchronization requirements and significantly boosts the achievable communication rate, it introduces a critical trade-off regarding energy efficiency. Unlike the standard on-off keying (OOK) modulation—where transmitting a logic `0' requires no molecular emission, thereby conserving energy—our array-based scheme mandates the generation and release of two distinct molecular pulses for every transmitted symbol. In the highly resource-constrained environment of intra-body nanonetworks, this increased energetic cost is non-trivial. In conventional time-slotted or rate-based encoding schemes, synchronization loss directly translates to catastrophic information loss. If the transmitter and receiver fall out of sync, transmitted pulses will be interpreted in the wrong time slots, drastically increasing the bit error rate and destroying the reliability of the communication link. In scenarios where autonomous nanomachines must accurately bridge severed neural pathways to transmit critical biological instructions (e.g., motor commands), such information loss renders the system entirely ineffective. Therefore, prioritizing synchronization-free transmission and higher channel capacity over raw energy expenditure is a highly justified design choice. Future implementations may explore adaptive encoding arrays or leverage localized biochemical energy-harvesting techniques to mitigate this energetic overhead.
	
	Furthermore, the physical and biological constraints of the synaptic cleft impose natural upper bounds on the practical transmission rate. Operating at the highest theoretical rates deduced from our numerical results could potentially outpace the biological clearance mechanisms, such as neurotransmitter reuptake by astrocytes or enzymatic degradation. If the emission rate of distinct molecules (e.g., Glutamate and GABA) exceeds the capacity of these clearing processes, residual molecules may lead to receptor saturation at the post-synaptic interface, thereby exacerbating inter-symbol interference. While our current analytical framework utilizes an additive Gamma noise model to encapsulate temporal propagation jitter effectively, it abstracts away the complex nonlinearities of receptor binding kinetics and resource depletion. Integrating these dynamic biological constraints into the capacity optimization problem represents a vital and promising direction for future comprehensive in vivo modeling.
	
	\section{Conclusion}
	In this paper, we proposed a novel neuro-spike array-based communication scheme designed to restore information transfer between damaged neurons utilizing auxiliary nano-machines. To overcome the stringent time-synchronization constraints of conventional neuro-spike models, our scheme encodes information through the sequential arrangement of distinct molecular pulses (e.g., excitatory and inhibitory neurotransmitters) emitted into the synaptic cleft. We analytically modeled the temporal jitter associated with the propagation of these pulses as an additive Gamma noise channel. Because the receiving nano-machines rely on specific target receptors to distinguish the pulses structurally, the necessity for exact time synchronization between transmitter and receiver is entirely eliminated.
	
	To rigorously evaluate the system performance, we derived closed-form theoretical expressions for the bit error probability and the probability of inter-symbol interference (ISI) occurrence. Subsequently, we obtained the achievable communication rate of the proposed scheme. Extensive numerical analysis demonstrated that our array-based architecture significantly outperforms traditional on-off keying (OOK) modulation operating over symbol-synchronized binary and Z-channel models. Specifically, the proposed scheme enhances the achievable communication rate by $75\,\% - 150\,\%$ across various diffusion coefficients, proving its robustness and high efficiency for future intra-body nanonetwork applications.
	
	\appendix
	
	\section{Proof of Theorem 1}
	The probability that symbol `0' is transmitted by the TN and correctly received by the RN is derived as:
	\begin{equation} \tag{A.1}
		{{\text{Pr}}}(x=0|y=0) = {{\text{Pr}}}({t_B} > {t_A} - {T_o}) = \left\{ \begin{array}{l}
			1,\,\,\,\,\,\,\,\,\,\,\,\,\,\,\,\,\,\,\, {t_A} \leq {T_o},\\
			\omega ({T_o}),\,\,\,\,\,\,\,\,{t_A} > {T_o}.
		\end{array} \right.
		\label{E17}
	\end{equation}
	We denote the random propagation delays of pulses $A$ and $B$ by $t_A$ and $t_B$, respectively.
	Since $t_A$ and $t_B$ share the identical PDF given in (\ref{E14}), i.e., $f_{\delta}(t) = f_{t_A}(t) = f_{t_B}(t)$, we can conclude that ${{\text{Pr}}}(x=0|y=0) = {{\text{Pr}}}(x=1|y=1)$. Consequently, we can rewrite the expression in (\ref{E17}) in a more compact form as:
	\begin{equation} \tag{A.2}
		{{\text{Pr}}}(x=0|y=0) = 1 \times {{\text{Pr}}}({t_A} \leq {T_o}) + \omega ({T_o}) \times {{\text{Pr}}}({t_A}>{T_o}).
		\label{E18}
	\end{equation}
	By employing the joint PDF of $t_A$ and $t_B$, the function $\omega ({T_o})$ is evaluated as:
	$$\omega ({T_o}) = \int\limits_{{T_o}}^\infty {\int\limits_{{t} - {T_o}}^\infty {f_{t_A\, t_B}({t},{t'})d{t'}d{t}} }=$$
	\begin{equation} \tag{A.3}
		\int\limits_{{T_o}}^\infty {\int\limits_{{t} - {T_o}}^\infty {{f_{{t_A}}}({t}){f_{{t_B}}}({t'})d{t'}d{t}} }= \int\limits_{{T_o}}^\infty {{f_{{t_A}}}({t})\frac{{\Gamma (\alpha ,\beta ({t} - {T_o}))}}{{\Gamma (\alpha )}}d{t}},
		\label{E19}
	\end{equation}
	where $\Gamma (\alpha,\beta t) = \int\limits_{\beta t}^\infty {{q^{\alpha - 1}}{e^{ - q}}dq}$ is the upper incomplete Gamma function, and $f_{t_A\,t_B}(t,t')$ represents the joint PDF of $t_A$ and $t_B$. We can substitute $f_{t_A\,t_B}({t},{t'}) = f_{t_A}({t})f_{t_B}({t'})$ since the pulse delays are assumed to be strictly independent. Furthermore, the probability ${{\text{Pr}}}({t_A} \leq {T_o})$, representing the scenario where $t_A$ does not exceed the intermission time $T_o$, is derived as:
	\begin{equation} \tag{A.4}
		{{\text{Pr}}}({t_A} \leq {T_o}) = \int\limits_0^{{T_o}} {{f_{{t_A}}}({t})d{t} = \frac{{\gamma (\alpha ,\beta {T_o})}}{{\Gamma (\alpha )}}}.
		\label{E20}
	\end{equation}
	By inserting (\ref{E19}) and (\ref{E20}) into (\ref{E18}), ${{\text{Pr}}}(x=0|y=0)$ is fully resolved as (\ref{E21}). \,\,\,\, $\blacksquare $
	
	\section{Proof of Theorem 2}
	We can formulate the probability of ISI occurrence, $P_{\text{ISI}}$, using the principle of complements as:
	$${P_{\text{ISI}}} = 1 - {P_{\text{NoISI}}} = 1 - {{\text{Pr}}}\left( {(\varphi ' > \varphi) \cap (\chi ' > \chi)} \right)$$
	\begin{equation} \tag{B.1}
		=1 - {{\text{Pr}}}\left( {\chi < \chi '|\varphi < \varphi '} \right){{\text{Pr}}}(\varphi < \varphi '),
		\label{E32}
	\end{equation}
	where $P_{\text{NoISI}}$ denotes the probability that a given symbol remains entirely free from interference caused by adjacent symbols. Additionally, the conditional probability ${{\text{Pr}}}\left( {\chi < \chi '|\varphi < \varphi '} \right)$ is obtained as:
	$${{\text{Pr}}}\left( {\chi < \chi '|\varphi < \varphi '} \right) ={{\text{Pr}}}\left( {{t'_A}+ T < \chi '|{{t'}_B} + {T_o} + T > \varphi } \right) \times {{\text{Pr}}}\left( {{{t'}_A} + T > {{t'}_B} + {T_o} + T} \right)$$
	\begin{equation} \tag{B.2}
		+ {{\text{Pr}}}\left( {{{t'}_B} + {T_o} + T < \chi '|{{t'}_A} + T > \varphi} \right) \times {{\text{Pr}}}\left( {{{t'}_B} + {T_o} + T > {{t'}_A} + T} \right).
		\label{E33}
	\end{equation}
	We then examine the two mutually exclusive cases defining $\varphi '$ and $\chi$: either $\chi = {t'_A} + T$ and $\varphi ' = {t'_A} + {T_o} + T$, or conversely, $\varphi ' = {t'_A} + T$ and $\chi = {t'_A} + {T_o} + T$. Owing to the independence of ${t'_A}$ and ${t'_B}$, the expression in (\ref{E33}) simplifies to:
	$${{\text{Pr}}}\left( {\chi < \chi '|\varphi < \varphi '} \right) = {{\text{Pr}}}\left( {{{t'}_A} < \chi ' - T} \right){{\text{Pr}}}\left( {{{t'}_A} > {{t'}_B} + {T_o}} \right)$$
	\begin{equation} \tag{B.3}
		+ {{\text{Pr}}}\left( {{{t'}_B} < \chi ' - {T_o} - T} \right){{\text{Pr}}}\left( {{{t'}_A} < {{t'}_B} + {T_o}} \right).
		\label{E34}
	\end{equation}
	Based on the geometric definitions from Fig. \ref{pic-7}, we substitute ${{\text{Pr}}}\left( {{{t'}_A} < {{t'}_B} + {T_o}} \right) = {{\text{Pr}}}(x=0|y=0)$ and its complement ${{\text{Pr}}}( {{{t'}_A} > {{t'}_B} + {T_o}} ) = 1 - {{\text{Pr}}}(x=0|y=0)$. Consequently, (\ref{E34}) can be rewritten as:
	$${{\text{Pr}}}(\chi < \chi '|\varphi < \varphi ') ={{\text{Pr}}}\left( {{{t'}_A} < \chi ' - T} \right) \times \left( {1 - {{\text{Pr}}}(x=0|y=0)} \right) $$
	\begin{equation} \tag{B.4}
		+ {{\text{Pr}}}\left( {{{t'}_B} < \chi ' - {T_o} - T} \right) \times {{\text{Pr}}}(x=0|y=0).
		\label{E35}
	\end{equation}
	Recognizing that ${{\text{Pr}}}\left( {{{t'}_A} < \chi ' - T} \right) = {{\text{Pr}}}\left( {{{t'}_A} - \chi ' < - T} \right)$, we introduce the substitution $s = - \chi '$. Utilizing ${f_{\chi '}}(t)$, the PDF of the auxiliary random variable $s$ is derived as:
	$${f_s}(t) = f_{\delta}( - t - 2T) \times \frac{{\Gamma (\alpha , - t - 2T - {T_o})}}{{\Gamma (\alpha )}}$$
	\begin{equation} \tag{B.5}
		+ f_{\delta}( - t - {T_o} - 2T) \times \frac{{\Gamma (\alpha , - t - 2T)}}{{\Gamma (\alpha )}},\,\,\,\,\,\,\,\,\,\,\, t \le - 2T.
		\label{E36}
	\end{equation}
	Because $t_A$ and $s$ are statistically independent, the PDF of their sum, $t_A+s$, is determined via convolution:
	\begin{equation} \tag{B.6}
		{f_{{t_A} + s}}(t) = \int\limits_m^\infty {f_{\delta}(v){f_s}(t - v)dv},
		\label{E37}
	\end{equation}
	where the lower bound of integration is $m = \max (0,t + 2T).$ Incorporating (\ref{E37}), the required probability evaluates to:
	\begin{equation} \tag{B.7}
		{{\text{Pr}}}\left( {{t'_A} < \chi' - T} \right) = {{\text{Pr}}}\left( {{{t'}_A} + s < - T} \right) = \int\limits_{ - \infty }^{ - T} {{f_{{t_A} + s}}(t)dt}.
		\label{E38}
	\end{equation}
	Following an analogous procedure, the probability ${{\text{Pr}}}\left( {{{t'}_A} < \chi ' - {T_o} - T} \right)$ from (\ref{E35}) is computed as:
	\begin{equation} \tag{B.8}
		{{\text{Pr}}}\left( {{{t'}_A} < \chi' - {T_o} - T} \right) = \int\limits_{ - \infty }^{ - {T_o} - T} {{f_{{t_A} + s}}(t)dt}.
		\label{E39}
	\end{equation}
	To simplify notation, we define the integral function $I(u) = \int\limits_{ - \infty }^{ - u} {{f_{ {t_A}+s}}(t)dt}$. Hence, the conditional probability reduces to:
	\begin{equation} \tag{B.9}
		{{\text{Pr}}}\left( {\chi < \chi'|\varphi < \varphi '} \right) = I(T)\left( {1 - {{\text{Pr}}}(x=0|y=0)} \right)+ I({T_o} + T){{\text{Pr}}}(x=0|y=0).
		\label{E40}
	\end{equation}
	Inserting (\ref{E40}) back into our initial formulation (\ref{E32}), we obtain:
	\begin{equation} \tag{B.10}
		{P_{\text{ISI}}} = 1 - ( {I(T) \left( {1 - {{\text{Pr}}}(x=0|y=0)} \right) }+{I({T_o} + T){{\text{Pr}}}(x=0|y=0)}){{\text{Pr}}}(\varphi' > \varphi).
		\label{E41}
	\end{equation}
	To evaluate the remaining term ${{\text{Pr}}}(\varphi' \leq \varphi)$, we introduce the transformation $r=-\varphi$. This yields:
	\begin{equation} \tag{B.11}
		{{\text{Pr}}}(\varphi' \leq \varphi) = {{\text{Pr}}}\left( {\varphi' + ( - \varphi) \leq 0} \right) = {{\text{Pr}}}(\varphi' + r \leq 0).
		\label{E42}
	\end{equation}
	From the previously determined ${f_\varphi}( t)$, the PDF of $r$ is:
	\begin{equation} \tag{B.12}
		{f_r}(t) = {f_\varphi}( - t) =f_{\delta}( - t)\gamma \left( {\alpha ,\beta \left( { - t - {T_o}} \right)} \right) + f_{\delta}( - t - {T_o})\gamma \left( {\alpha ,\beta \left( { - t} \right)} \right), \,\,\,\,\,\,\,\,\,\,\, t \le - {T_o}.
		\label{E43}
	\end{equation}
	Assuming independence between $\varphi'$ and $r$, the PDF of the combined variable $\varphi'+r$ is established through convolution:
	\begin{equation} \tag{B.13}
		{f_{\varphi' + r}}(t) = \int\limits_{ - \infty }^\infty {{f_{\varphi'}}(v){f_r}(t - v)dv = \int\limits_T^\infty {{f_{\varphi'}}(v){f_r}(t - v)dv} }= \int\limits_n^\infty {{f_{\varphi'}}(v){f_r}(t - v)dv},
		\label{E44}
	\end{equation}
	where the integration limit is $n = \max (t + {T_o},T)$. Employing this PDF, we calculate:
	\begin{equation} \tag{B.14}
		{{\text{Pr}}}(\varphi' \leq \varphi) = {{\text{Pr}}}(\varphi' + r \leq 0) = \int\limits_{ - \infty }^0 {{f_{\varphi' + r}}(t)dt}.
		\label{E45}
	\end{equation}
	Let us assign the constant $K$ to represent this integral, such that $K = \int\limits_{ - \infty }^0 {{f_{\varphi' + r}}(t)dt} $. Therefore, the complementary probability becomes ${{\text{Pr}}}(\varphi < \varphi') = 1 - K$.
	Substituting this result into (\ref{E41}), the total ISI probability is expressed as:
	\begin{equation} \tag{B.15}
		{P_{\text{ISI}}} = 1 - ( {I(T)\left( {1 - {{\text{Pr}}}(x=0|y=0)} \right)}+ {I(T + {T_o}){{\text{Pr}}}(x=0|y=0)}) \times (1 - K).
		\label{E47}
	\end{equation}
	Expanding this equation by substituting the definitions for ${\text{Pr}}(x=0|y=0)$, $I(T)$, and $I(T_o+T)$ confirms the final expression presented in (\ref{E48}). \,\,\,\,$\blacksquare $

	\section{Proof of Theorem 3}
	As observed in (\ref{E48}), the probability of ISI occurrence, $P_{\text{ISI}}$, depends on the successful transmission probability, i.e., ${\text{Pr}}(x=0|y=0)$. Consequently, the overall error probability of the array-based scheme, denoted as $P_e$, is derived as:
	\begin{equation} \tag{C.1}
		{P_e} = 1 - {P_{\text{NoISI}|(x=0|y=0)}}{{\text{Pr}}}(x=0|y=0),
		\label{E49}
	\end{equation}
	where $P_{\text{NoISI}|(x=0|y=0)}$ represents the conditional probability that a symbol does not interfere with adjacent symbols given that it was correctly transmitted. According to the delay variables established in Fig. \ref{pic-7}, this probability is obtained as:
	$$P_{\text{NoISI}|(x=0|y=0)} = {{\text{Pr}}}\left( {(\varphi' > \varphi) \cap (\chi < \chi ')|(x=0|y=0)} \right)$$
	\begin{equation} \tag{C.2}
		= {{\text{Pr}}}\left( {\chi < \chi '|\varphi < \varphi',(x=0|y=0)} \right)\times{{\text{Pr}}}\left( {\varphi < \varphi'|(x=0|y=0)} \right).
		\label{E50}
	\end{equation}
	The first term, ${{\text{Pr}}}\left( {\chi < \chi'|\varphi < \varphi',(x=0|y=0)} \right)$, is evaluated as:
	$${{\text{Pr}}}\left( {\chi < \chi'|\varphi < \varphi',(x=0|y=0)} \right)= {{\text{Pr}}}\left( {{{t'}_B} + {T_o} + T < \chi'|{{t'}_A} + T > \varphi} \right)$$
	\begin{equation} \tag{C.3}
		= {{\text{Pr}}}\left( {{{t'}_B} < \chi' - {T_o} - T} \right) = I({T_o} + T),
		\label{E51}
	\end{equation}
	where the result ${{\text{Pr}}}\left( {{t'_B} < \chi' - {T_o} - T} \right) = I({T_o} + T)$ was previously established in (\ref{E39}). Furthermore, the second term, ${{\text{Pr}}}\left( {\varphi < \varphi'|(x=0|y=0)} \right)$, is obtained as:
	\begin{equation} \tag{C.4}
		{{\text{Pr}}}\left( {\varphi < \varphi'|(x=0|y=0)} \right) = {{\text{Pr}}}\left( {{t'_A} + T > \varphi} \right)= {{\text{Pr}}}\left( {{t'_A} - \varphi > - T} \right).
		\label{E52}
	\end{equation}
	By applying the substitution $r=-\varphi$, we rewrite (\ref{E52}) as:
	\begin{equation} \tag{C.5}
		{{\text{Pr}}}\left( {\varphi < \varphi'|(x=0|y=0)} \right) = {{\text{Pr}}}\left( {{t'_A} - \varphi > - T} \right)= {{\text{Pr}}}\left( {{t'_A} + r > - T} \right).
		\label{E53}
	\end{equation}
	The PDF of $r$, denoted by $f_r(t)$, was derived in (\ref{E43}). Because $r$ and ${{t'_A}}$ are statistically independent, the PDF of their sum, ${{t'_A}}+r$, is derived via convolution:
	\begin{equation} \tag{C.6}
		{f_{{t'_A} + r}}(t) = \int\limits_{ - \infty }^{ + \infty } {f_{\delta}(v){f_r}(t - v)dv } = \int\limits_0^{ + \infty } {f_{\delta}(v){f_r}(t - v)dv} = \int\limits_h^{ + \infty } {f_{\delta}(v){f_r}(t - v)dv},
		\label{E54}
	\end{equation}
	where the lower limit is $h = \max \left( {0,t + {T_o}} \right)$. Since the variables are identically distributed, ${f_{{t_A} + r}}(t) = {f_{{{t'_A}} + r}}(t)$. Using ${f_{{t_A} + r}}(t)$, the conditional probability ${{\text{Pr}}}\left( {\varphi < \varphi'|\left( {x=0|y=0} \right)} \right)$ is evaluated as:
	\begin{equation}
		{{\text{Pr}}}\left( {\varphi < \varphi'|\left( {x=0|y=0} \right)} \right) = {{\text{Pr}}}\left( {{t'_A} + r > - T} \right) =
		\nonumber
	\end{equation}
	\begin{equation} \tag{C.7}
		1 - {F_{{t_A} + r}}( - T) =1 - \int\limits_{ - \infty }^{ - T} {{f_{{t_A} + r}}(t)dt}.
		\label{E55}
	\end{equation}
	We define the integral $J(u) = \int\limits_{ - \infty }^{ - u} {{f_{{t_A} + r}}(t)dt}$, simplifying the expression to ${{\text{Pr}}}\left( {\varphi < \varphi'|(x=0|y=0)} \right) = 1 - J(T)$.
	Inserting the expressions from (\ref{E51}) and (\ref{E55}) into (\ref{E50}), $P_{\text{NoISI}|(x=0|y=0)}$ is fully derived as:
	$$P_{\text{NoISI}|(x=0|y=0)} = {{\text{Pr}}}\left( {\chi < \chi '|\varphi < \varphi',(x=0|y=0)} \right)\times$$
	\begin{equation} \tag{C.8}
		{{\text{Pr}}}\left( {\varphi < \varphi'|(x=0|y=0)} \right)= I({T_o} + T)\left( {1 - J(T)} \right).
		\label{E57}
	\end{equation}
	Finally, substituting ${P_{\text{NoISI}|(x=0|y=0)}}$ into (\ref{E49}) yields the complete error probability $P_e$ as presented in (\ref{E58}). \,\,\,\,$\blacksquare $

	\section{Rate of On-Off Keying Scheme in the Z-Channel and Binary Channel Models}
	In this appendix, we detail the standard on-off keying (OOK) scheme operating over the Z-channel and binary channel models, which serve as baselines for our performance evaluation. In the OOK scheme within the Z-channel model \cite{arifler2011capacity}, a single molecular pulse is emitted by the TN at the beginning of the time slot to represent logic `1'. If this pulse reaches the RN within the designated time slot, $T$, the RN successfully detects the logic bit `1' with probability $\eta$. Otherwise, the RN erroneously detects a logic bit `0' with probability $1-\eta$. Therefore, the success probability $\eta$ is obtained as:
	$$\eta = \int\limits_0^T {f_{\delta}(t)dt = \frac{{\gamma (\alpha ,\beta T )}}{{\Gamma (\alpha )}}}.$$
	To transmit bit `0', the TN emits no pulse during the time slot. In the idealized Z-channel model, the transmission of bit `0' is assumed to be perfectly successful. By exploiting $\eta$, the transition matrix for the Z-channel, denoted by ${P_z}(y|x)$, is formulated as:
	$$
	{P_z}(y|x) = \left( {\begin{array}{*{20}{c}}
			1&0\\
			{1 - \eta }&\eta
	\end{array}} \right),
	$$
	where $x$ and $y$ are the channel input and output bits, respectively. Based on ${P_z}(y|x)$, the capacity of the Z-channel model, $C_z$, is derived as:
	\begin{equation} \tag{D.1}
		{C_z} = \mathop {\max }\limits_{Q(x)} I(X;Y) = \mathop {\max }\limits_{Q(x)} \sum\limits_{x \in X} {\sum\limits_{y \in Y} {Q(x){P_z}(y|x){{\log }_2}} \left( {\frac{{{P_z}(y|x)}}{{\sum\limits_{x' \in X} {{P_z}(y|x'} )}}} \right)},
		\label{E63}
	\end{equation}
	where $I(X;Y)$ denotes the mutual information between the channel's input and output. The input probability distribution is defined as $Q(1) = {{\text{Pr}}}(x = 1)$ and $Q(0) = {{\text{Pr}}}(x = 0)$. Since $C_z$ represents the capacity in bits per transmission, the corresponding communication rate in the time domain is given by ${C_z}/{T}$ (bps).
	
	\begin{figure}[H]
		\centering
		\includegraphics[width=.8\textwidth]{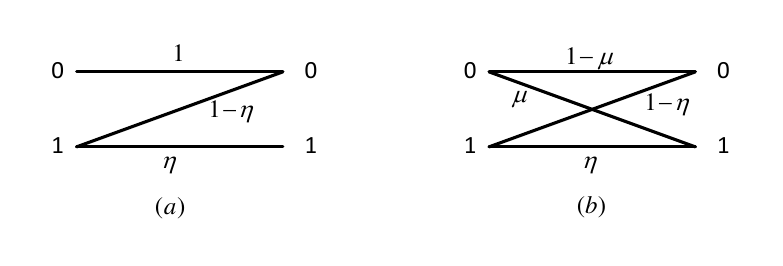}
		\caption{(a) Z-channel model. (b) Binary channel model.}
		\label{pic-99}
	\end{figure}
	
	While the Z-channel model assumes that all transmissions of bit `0' are flawlessly received, this represents an oversimplification in molecular environments. In reality, molecular pulses emitted in preceding intervals may experience extended diffusion latency (the heavy tail of the Gamma distribution) and arrive at the RN during subsequent time slots, causing a false positive. To account for these erroneous detections of logic bit `0', the binary channel model is utilized, as depicted in Fig. \ref{pic-99} and introduced in \cite{mahfuz2010characterization}.
	
	Assuming that the current transmission of bit `0' is solely affected by a bit `1' transmitted in the immediately preceding interval, the probability that a bit `0' is falsely detected as a `1' by the RN, denoted as $\mu$, is calculated by:
	\begin{equation} \tag{D.2}
		\mu = Q(1)\int\limits_T ^{2 T } {f_{\delta}(t)dt} = Q(1) \times \frac{{\gamma (\alpha ,2\beta T) - \gamma (\alpha ,\beta T )}}{{\Gamma (\alpha )}}.
		\label{E64}
	\end{equation}
	By incorporating both $\mu$ and $\eta$, the transition matrix for the more realistic binary channel model, ${P_b}(y|x)$, is described as:
	$$
	{P_b}(y|x) = \left( {\begin{array}{*{20}{c}}
			1-\mu&{\mu} \\
			{1 - \eta }&\eta
	\end{array}} \right).
	$$
	Using $P_b(y|x)$, the achievable rate of the binary channel model in bits per transmission, $C_b$, is obtained by solving the mutual information optimization problem described in (\ref{E63}), akin to the Z-channel approach. Finally, the communication rate for the binary channel model is evaluated as ${C_b}/{T}$ (bps).
	
	\bibliographystyle{IEEEtran}
	\bibliography{ref2}
\end{document}